\documentclass[extra,mreferee]{gji} 
\usepackage{timet}
\usepackage{amsmath,amssymb,amsfonts,latexsym}[mathlines] 
\usepackage{mathrsfs}
\usepackage{graphicx}
\usepackage{changepage}
\usepackage{color}
\usepackage{lineno}
\usepackage[finalnew]{trackchanges} 
\usepackage{soul}

\title[
Rendering potency beachballs into fault shape
]
  {
  Smooth surface reconstruction of earthquake faults from distributed moment-potency-tensor solutions
}
\author[D. Sato, Y. Yagi, R. Okuwaki, \& Y. Fukahata
]
  {Dye SK Sato$^{1}$, Yuji Yagi$^{2}$, Ryo Okuwaki$^{2}$, \& Yukitoshi Fukahata$^{3}$
\\
  $^1$ Research Institute for Marine Geodynamics, Japan Agency for Marine-Earth Science and Technology, \\\,\,\,\,Yokohama, Kanagawa 236-0001, Japan
\\
  $^2$ Institute of Life and Environmental Sciences, University of Tsukuba, Tsukuba, Ibaraki 305-8572, Japan\\
  $^3$ Disaster Prevention Research Institute, Kyoto University, Uji, Kyoto 611-0011, Japan
  }
\date{Received 20xx Xxxx xx; in original form 20xx Xxxx xx}
\pagerange{\pageref{firstpage}--\pageref{lastpage}}
\volume{xxx}
\pubyear{xxxx}

\begin{document}

\label{firstpage}

\maketitle


\begin{summary}
The earthquake fault as observed by seismic motion primarily manifests as a surface of displacement discontinuity within a linear elastic continuum. The displacement discontinuity and the surface normal vector ($n$-vector) of this idealized earthquake source are measured by the tensor of potency, which is seismic moment normalized by stiffness. We exploit this theoretical relation to formulate an inverse problem of reconstructing a smooth, three-dimensional fault surface from an areal density field of the potency tensor. 
In this problem, the surface is represented by an elevation field that parametrizes the vertical variation of the surface relative to a reference, and the nodal planes of a given potency-density-tensor field describe the $n$-vector field.
The remaining subject is the $n$-vector-to-elevation transform, the operation inverse to defining the $n$-vector field on a given surface. 
Whereas this transform is a well-posed one-to-one mapping in two dimensions where the $n$-vector has one degree of freedom, the transform becomes overdetermined in three dimensions because the $n$-vector has two degrees of freedom while the scalar elevation has only one, generally admitting no solution. 
This overdetermination originates from a reduction in degrees of freedom from six to five upon modeling the source as a displacement discontinuity rather than general potency density, namely inelastic strain. The sixth degree of freedom unmodeled by displacement discontinuities and $n$-vectors manifests as a local violation of the determinant-free constraint in point potency sources; however, in areal sources of potency density, it raises a conflict with the global consistency of the $n$-vector field.
Recognizing that this conflict derives from the capacity of the potency-density-tensor field to describe inelastic strain source incompatible with displacement discontinuity on a surface, we explicitly introduce an a priori constraint to define the fault surface as the smooth surface that best approximates the surface distribution of inelastic strain by displacement discontinuity.
We derive an analytical solution for the surface reconstruction thus formulated and demonstrate its ability to reproduce smooth three-dimensional surfaces from synthetic noisy $n$-vector fields. 
Lastly, we integrate the derived formula into the potency density tensor inversion and validate it in an application to the 2013 Balochistan earthquake. 
The estimated fault geometry agrees better with the observed fault trace than that of the previously proposed quasi-two-dimensional surface reconstruction, highlighting the importance of accounting for three-dimensional fault geometry.
\end{summary}


\section{Preliminaries}
Geometry of earthquake faults predetermines probable rupture scenarios, governing the key phases of initiation, termination, stagnation, and branching~\citep{das1977fault,aki1979characterization,kame2003effects,kase2006spontaneous,wesnousky2006predicting,bruhat2016rupture,ozawa2023quantifying}. 
Fault geometry inference from observed data thus provides critical insights into the complex source process that occurred~\citep{barka1988strike,kodaira2006cause,ando2018dynamic,howarth2021spatiotemporal}. 
The endeavor of fault reconstruction dates back to the global centroid-moment-tensor (CMT) project~\citep{dziewonski1981determination}. The earthquake fault as a point source has been described by the spatially integrated tensor product of the slip vector and the fault normal vector, called the potency tensor~\citep{ben1981seismic}, which is measured as the seismic-moment tensor~\citep{kanamori1975theoretical} normalized by the stiffness tensor. Resolving earthquake faults beyond this point-source description remains an ongoing challenge~\citep{matsu1987maximum,yabuki1992geodetic,volkov2017reconstruction,dutta2021simultaneous}. Distributed source inversions involving fault geometry inferences have been proved to have solutions~\citep{beretta2008determination,volkov2017reconstruction,aspri2020analysis,diao2023dislocations}, but how to solve them is under debate~\citep{ragon2018accounting,dutta2021simultaneous,shimizu2021construction,wei2023bayesian,aspri2025shape}.

A series of developments are based on slip inversion~\citep{chinnery1964strength}, which estimates an earthquake source as a displacement discontinuity on a boundary~\citep[or its gradient, dislocation;][]{matsu1977inversionI,matsu1977inversionII}. 
Slip inversion typically confines the model-parameter space to shear slip~\citep{chinnery1964strength}, while variants exist that estimate tensile faults as well~\citep{massonnet1998radar}. Either form of analysis could be termed slip-and-opening inversion or displacement-discontinuity inversion; however, as no definitive nomenclature has been established, this paper collectively refers to all inverse problems that model the earthquake source as a displacement discontinuity as `slip inversion' in a broad sense.
Within the framework of slip inversion, slip (and opening) distributions on faults are model parameters, and fault geometry is an additional set of parameters~\citep{savage1966surface,matsu1987maximum,fukahata2008non,ragon2018accounting} that determine the structure of the inference model, associated with hyperparameters in statistics~\citep{matsu1977inversionI}. 
The simplest, highly successful example along this line is configuration estimation using a rectangular fault, which infers the center of masses, angles, and lengths of a fault, besides the vector of the displacement discontinuity~\citep{savage1966surface,matsu1987maximum}. 
Detailed slip patterns can be estimated from a distributed slip inversion on a prescribed fault surface~\citep{yabuki1992geodetic}.
When nonplanar geometry is inferred in slip inversion, fault subdivisions with a finite number of planar subfaults have been widely utilized~\citep{matsu1977inversionI,matsu1977inversionII,loveless2010geodetic,avouac20142013}. 
This approach, however, suffers from the artificial stress singularity arising at subfault edges~\citep{romanet2020curvature,romanet2024mechanics,mallick2025smooth}. A primary objective of fault geometry inference is the estimation of the macroscopic fault geometry, which is coarse-grained as a smooth surface consistent with linear continuum mechanics~\citep{matsu2019physical,dutta2021simultaneous}. 
 
In slip inversion, where the estimation of displacement discontinuity is a linear problem, the estimation of fault geometry alters the problem into a nonlinear inference since the kernel function (Green's function) to relate slip and observed data depends on the geometry~\citep{yabuki1992geodetic}. This nonlinearity contrasts with the CMT analysis, where far-field conditions enabling the point-source approximation ensure the positional insensitivity of Green's functions, allowing for stable perturbative updates from a coarse initial solution~\citep{dziewonski1981determination}. 
The radiation pattern of Green's function amplitudes, which conveys source geometry information to the far-field, is determined by the potency representing the slip vector and the fault normal, rendering the CMT analysis an almost linearly-solvable inversion. In turn, this quasilinearity holds without the point-source approximation if the source is formulated as a density field of such potency. This realization paved the way for the potency density tensor inversion~\citep[PDTI,][]{kikuchi1991inversion,kasahara2010complex,shimizu2020development}, where the earthquake fault is expressed as a potency-density-tensor field distributed over a domain of negligible volume, a pseudo boundary.

In the PDTI, the model parameters correspond to the areal density field of the potency tensor set along a pseudo boundary. 
The displacement discontinuity and the fault normal are equivalently expressed by the nodal planes of the estimated potency density tensors that smoothly rotate over space and time.
By permitting potency incompatible with the normal of the pseudo boundary placed as a first-order approximation of the actual fault, the PDTI extracts geometry information consistent with observed seismic motions. 
Because the seismic moment density represents the deviation of stress from elasticity~\citep[stress glut,][]{backus1976momentI,backus1976momentII}, the seismic moment density normalized by the stiffness, the potency density describes the inelastic strain~\citep{kostrov1974seismic}. Consequently, the potency density tensor possesses six degrees of freedom as long as it is treated as a representation of inelastic strain within a finite volume, although it is essentially equivalent to the tensor product of a displacement discontinuity (slip and opening) and a fault normal in the limit of infinitesimal volumes; recall that the strain tensor has six degrees of freedom whereas the set of fault normal and displacement discontinuity has five degrees of freedom due to the normalization condition of the $n$-vector. 
Although counterintuitive, the inelastic strain source, the potency density, is intrinsically higher-dimensional than displacement discontinuity. Therefore, the PDTI can express multiple earthquake faults, such as a conjugate fault system and a set of disconnected faults, on a single model plane~\citep{yamashita2021consecutive,okuwaki2022oblique}, enabling the modeling of proximal faults even when the inter-surface distance falls below the observation limit~\citep{volkov2025stability} of discernibility.

The pioneering research of \citet{kikuchi1991inversion} first inferred a seismic source as inelastic strain, namely the potency density, distributed along a pseudo boundary instead of as displacement discontinuity across a volume-less interface. 
Nonetheless, their original PDTI is almost indistinguishable from planar fault subdivision and similar to multiple point source approaches~\citep{tsai2005multiple,yue2020resolving} because the prior constraint they employ is only the double-couple condition imposed as a traceless and determinant-free constraint, where the determinant-free constraint functions as the point-source displacement-discontinuity condition while the traceless condition eliminates the volumetric deformation. 
A major advantage of the PDTI stems from the affinity for incorporating a prior constraint upon the fault geometry, expressible as the hyperprior imposed onto geometrical hyperparameters in the framework of slip inversion, which is, to our knowledge, first introduced by \citet{kasahara2010complex}. 
The prior for the potency operates on both the displacement discontinuity and the $n$-vector field, and thus, smoothly varying macroscopic rotations of focal mechanisms are extracted by simply using a smoothness constraint on the model parameter in the PDTI. 
\citet{kasahara2010complex} have further developed a stable scheme by incorporating the propagation of Green's function errors~\citep{yagi2010waveform,yagi2011introduction} typified by structural uncertainties, which induce apparent rotations of radiation patterns for near-vertical take-off angles~\citep{ford2012event,chiang2014source} and travel-time perturbations~\citep{hu2023seismic,hu2025bayesian}. 
Whereas the original intention of \citet{kasahara2010complex} was rather to extract spatial variations in slip vectors and focal mechanisms themselves, 
\citet{shimizu2020development} point out that spatial patterns of the potency focal mechanism also document fault geometry as fault normal vector fields. 
The PDTI untangles source complexities in the perspective of fault geometry thereafter~\citep{okuwaki2020inchworm,tadapansawut2021rupture,fan2022fast,ohara2023complex,yagi2024barrier,inoue2025a}. 

The success of the PDTI stimulates a growing demand for earthquake fault reconstruction from the distributed potency solutions, namely reconstructing the spatial configuration of nonplanar faults from the normal-vector ($n$-vector) field indicated by the nodal planes of the estimated potency-density-tensor field~\citep{shimizu2021construction}. 
This task is analogous to reconstructing a surface from point-clouds of microseismicity or aftershocks~\citep{yukutake2011fluid,fuenzalida2013high,ross20203d,wang2021aftershock,sawaki2025fault}, 
but the input here is more sensitive to local fault orientations than to local fault positions. 
The problem to be solved is to convert the $n$-vector field $n(x,y)$ on the model $xy$-plane(s) into the elevation $z(x,y)$ of the nonplanar fault relative to the model plane(s). This transformation corresponds to 
an operation inverse to the $n$-vector's definition that assigns the normal $n$ on a surface elevated by $z$. 
What complicates this $n \mapsto z$ operation is the fact that the $n$-vector possesses two degrees of freedom whereas the scalar elevation $z$ has only a single degree of freedom. Except for special sets of the potency density tensors, it does not have a solution. 
After all, it is impossible to fully map the potency density that represents inelastic strain within a finite volume onto a zero-volume displacement discontinuity, a counterpart issue of the non-zero determinant not expressed by the displacement discontinuity encountered in point-source moment tensor solutions.

To construct fault geometry from the potency while preserving the framework of the PDTI that allows for the inelastic strain source beyond displacement discontinuity, a criterion becomes necessary to filter out potency components impermissible as displacement discontinuity in constructing a fault surface. \citet{shimizu2021construction} adopted an assumption of two-dimensional geometry, such as along-strike strike-angle variations, thereby reducing the $n$-vector field to a scalar field; spatially integrating estimated fault slopes given by the 90-degree rotations of $n$-vectors provides the estimated elevation in two dimensions. 
This study derives a criterion for reconstructing three-dimensional fault geometry for the case where the potency-density-tensor field constitutes a displacement-discontinuity field on a single smooth surface, the standard hypothesis of slip inversion. 
Through the formalization of the reconstruction problem, this work exposes the overdetermined, criterion-based nature of inferring fault geometry from the estimated potency, which compels us to acknowledge that the seismic source is not limited to displacement discontinuity: a dislocation-theoretic description of earthquake faults is not inherent but partly results from a subjective constraint imposed on the potency. We formulate earthquake fault reconstruction as an implementation of such a prior constraint. 

Consequently, this formalization reveals that the seismic source representation undergoes a fundamental dimensionality reduction when transitioning from general inelastic strain to displacement discontinuity. While a moment tensor possesses six degrees of freedom, its representation as a displacement discontinuity is restricted to five at each coordinate, which is an inevitable consequence of the compatibility in continuum mechanics. For a point source, it is a simple rank-2 approximation implemented by the determinant-free constraint. In contrast, for an areal source, this dimensionality reduction within the source description is the origin of the overdetermination in earthquake fault reconstruction.

\section{Model setting of earthquake fault reconstruction from distributed potency}
\subsection{Observation equations}
\label{subsec:forward}
Source geometry inference is based on the following three model equations.

The first model equation is the representation theorem for linear continua. 
The displacement $u_i$ of the $i$-th data component is expressed by linear combinations of 
potency areal density $D(\boldsymbol\xi)$ at coordinates $\boldsymbol\xi$ on faults $\Gamma$~\citep{kikuchi1991inversion}:
\begin{equation}
    u_i=\int_{\Gamma} d\Sigma(\boldsymbol\xi) G_i(\boldsymbol\xi) D(\boldsymbol\xi),
    \label{eq:RepTheorem}
\end{equation}
where $G_i(\boldsymbol\xi)$ denotes Green's function relating $u_i$ and $D(\boldsymbol\xi)$. 
The same formulation of geometry inference is applicable to time-dependent problems, and we incorporate the time dependence of $u$ and $D$ in the application to teleseismic data. 
The seismic moment is not uniquely defined across a stiffness discontinuity (e.g., a plate boundary) while the potency is well-defined even across it~\citep{ampuero2005ambiguity}. The observed surface displacement reflects the potency rather than the seismic moment, which requires additional information on the highly ambiguous fault stiffness~\citep{heaton1989static,matsu2019physical}. 
Against this background, we use potency to describe earthquake faults. 

The second model equation is the transformation of double-couple sources to fault slips. 
When the potency-density tensor $D$ represents a double-couple source, $D$ is expressed by the symmetrized tensor product of a slip vector ${\bf s}$ and a fault normal ${\bf n}$: 
\begin{equation}
    D= \frac{1}{2}({\bf s}{\bf n}^{\rm T}+{\bf n}{\bf s}^{\rm T}),
    \label{eq:defofnvector_PDTI}
\end{equation}
where the superscript ${\rm T}$ denotes the transpose. It should be emphasized that the potency density $D$ defined in eq.~(\ref{eq:defofnvector_PDTI}) is not restricted to shear slip or double-couple sources. Rather, eq.~(\ref{eq:defofnvector_PDTI}) represents an absolute requirement for any seismic source modeled as a displacement discontinuity $\bf{s}$ across a surface orthogonal to the orientation vector $\bf{n}$. Hereafter, we use the term `slip' as an intuitive shorthand for displacement discontinuity, consistent with our usage of `slip inversion' in a broad sense. 
Shear deformation corresponds to a traceless $D$, whereas opening deformation corresponds to a rank-one $D$ (s.t. $D = |D|\bf{n}\bf{n}^{\rm T}$). In the former case, eq.~(\ref{eq:defofnvector_PDTI}) indicates that an $n$-vector corresponds to one of the two nodal planes of potency in the sense of the beachball diagram for focal mechanisms. 
Equivalently, the $n$-vector is one of the four normalized sums of the eigenvectors of $D$ associated with the maximum and minimum eigenvalues, where the fourfold ambiguity arises from the signs of the eigenvectors. 
In the latter case, the $n$-vector is identified as the eigenvector corresponding to the eigenvalue with the largest absolute magnitude. Although this study primarily focuses on shear-dominant faults, the PDTI remains applicable to tensile faults.

The source inversion via eq.~(\ref{eq:RepTheorem}) and 
the nodal-plane extraction by eq.~(\ref{eq:defofnvector_PDTI}) have established methods of analysis. 
The potency density tensor inversion of eq.~(\ref{eq:RepTheorem}) is similar to the slip inversion~\citep{yabuki1992geodetic,yagi2011introduction}. 
The nodal-plane extraction for eq.~(\ref{eq:defofnvector_PDTI}) is the same as that for moment tensor solutions. 

The third equation is the correspondence between the $n$-vector field and the shape of a surface, the definitional identity of the $n$-vector: 
\begin{equation}
    {\bf n}(\boldsymbol\xi)\cdot
    d \boldsymbol\xi(\boldsymbol\xi)=0,
    \label{eq:defofnvector}
\end{equation}
which states that the normal vector ${\bf n}(\boldsymbol\xi)$ at a coordinate $\boldsymbol\xi$ on a surface $\Gamma$ is orthogonal to the infinitesimal shift $d \boldsymbol\xi$ along $\Gamma$ at $\boldsymbol\xi$. 
A `smooth surface' in this paper denotes a surface whose shape is defined by eq.~(\ref{eq:defofnvector}), termed a `regular surface' in differential geometry. 
Equation~(\ref{eq:defofnvector}) describes the condition under which an infinitesimal region at coordinate $\boldsymbol \xi$ is expressed as a tangent plane.
The orthogonality to the tangent plane defines the normal vector of the nonplanar surface, and a relative location between two arbitrary points on a smooth surface is given by the line integral $\int d\boldsymbol\xi$ that satisfies eq.~(\ref{eq:defofnvector}). 
Since the $n$-vector field is equivalent to the shape of the smooth surface up to a rigid-body translation, the $n$-vector field fully encodes the macroscopic smooth shape of a fault. 

In the following analysis, we assume for simplicity that the $n$-vector expected from a given potency $D$ corresponds to the nodal plane of $D$ closest to a known reference normal ${\bf n}^{\circ}$ (e.g., the expectation from the first-motion solution). That is, the angular deviation of the $n$-vector from the reference is assumed to be within 45 degrees. 
Here, we implicitly suppose the case where a coarse estimate of the $n$-vector field is available within a 45$^\circ$ error margin; the orientation of ${\bf n}^{\circ}$ can be spatially varying if necessary. 
Similarly, we assume that a single location $\boldsymbol\xi^\circ$ on (or at least adjacent to) the fault (e.g., the rupture initiation point and a point along the observed fault trace) is known, and its estimation errors are neglected. 
Equation~(\ref{eq:defofnvector}) only describes the relative locations between the given points on a surface, and the absolute location is shifted by $\boldsymbol\xi^\circ$. 
The two quantities $\boldsymbol\xi^\circ$ and ${\bf n}^{\circ}$ need not reflect an identical point-source representation.

In the PDTI, $D$ is defined via eq.~(\ref{eq:RepTheorem}) as a source representation that generally possesses six degrees of freedom (DOFs) relative to observations, implying that eq.~(\ref{eq:defofnvector_PDTI}) is not strictly satisfied. A symmetric matrix composed of two vectors, as in eq.~(\ref{eq:defofnvector_PDTI}), has a rank of at most two and is subject to a determinant-free constraint, meaning that a displacement-discontinuity source cannot exceed five DOFs. Indeed, the combined DOFs of a displacement discontinuity and an $n$-vector amount to five. Equation (\ref{eq:defofnvector_PDTI}) restricts the direction of the opening mode to the fault-normal by imposing vanishing tangential-tangential components. 
Although a traceless condition is typically imposed to exclude tensile components when assuming pure shear slip, the tensile component by default serves to resolve the degeneracy between the fault normal and the displacement discontinuity within the moment/potency tensor.

Equation (\ref{eq:defofnvector_PDTI}) is derived as an integrability condition, arising from the compatibility condition in continuum mechanics, for the surface distribution of inelastic strain (Appendix~\ref{app:compatibility}). There is thus no physical requirement for it to be satisfied as long as the source is treated as inelastic strain within a finite volume. The excess degrees of freedom that cannot be represented by a surface capture the volumetric information.
Therefore, eq.~(\ref{eq:RepTheorem}) should not be read as a mere restatement of the representation theorem for displacement discontinuities. It is instead a representation theorem for finite-volume sources where thickness is neglected. As demonstrated in the following section, the primary objective in geometry reconstruction from the potency is to extract the surface normal $n$ (eq.~\ref{eq:defofnvector}), defined in the displacement-discontinuity limit, from a source that is not strictly an ideal surface. For a point source, this task reduces to ignoring the eigenvalue with the smallest absolute magnitude to satisfy the determinant-free condition. For an areal source, it requires addressing the global constraints of eq.~(\ref{eq:defofnvector}), which act through infinitesimal motions along the surface.

\subsection{Decoupling surface reconstruction from potency estimation}
\begin{figure*}
 \includegraphics[width=130mm]{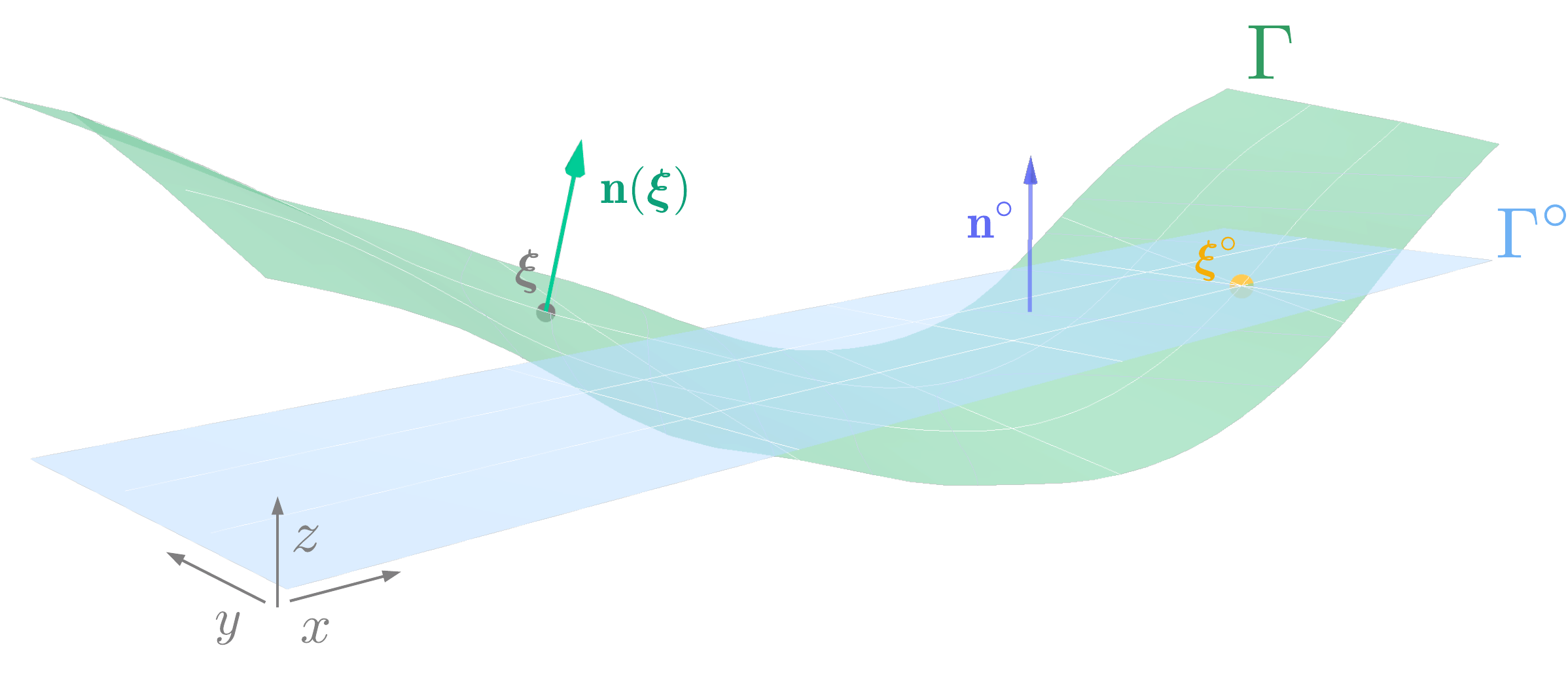}
 \caption{
 Parametrization of a fault surface $\Gamma$ using a reference surface $\Gamma^\circ$. 
 The normal vector ${\bf n}$ of $\Gamma$ depends on the location $\boldsymbol\xi$ parametrized as $(x,y,z)$. 
 The $x$-$y$ plane describes the coordinate of $\boldsymbol\xi$ along $\Gamma^\circ$, and the $z$ axis represents the vertical elevation of $\Gamma$ relative to $\Gamma^\circ$. 
 In the schematic, $\Gamma^\circ$ is drawn as a flat surface being perpendicular to a reference vector ${\bf n}^\circ$ and passing through a point $\boldsymbol\xi^\circ$ on $\Gamma$. 
}
 \label{fig:1}
\end{figure*}

The PDTI evaluates the potency density $D$ and the associated $n$-vector field through the well-known equations (\ref{eq:RepTheorem}) and (\ref{eq:defofnvector_PDTI}), and the $n$-vector field is translated into $\Gamma$ through eq.~(\ref{eq:defofnvector}). 
Here is a problematic circular dependency between $D$ and $\Gamma$: the domain on which the potency density $D$ is defined is the fault surface $\Gamma$ (eq.~\ref{eq:RepTheorem}), which is yet to be evaluated by $D$ (eq.~\ref{eq:defofnvector}). 
To infer $D$ prior to reconstructing $\Gamma$, 
the representation theorem (eq.~\ref{eq:RepTheorem}) must be projected onto a fixed surface, here termed the reference surface, denoted by $\Gamma^\circ$ (Fig.~\ref{fig:1}).

In this study, $\Gamma^\circ$ is a plane perpendicular to ${\bf n}^\circ$ passing through $\boldsymbol\xi^\circ$ (Fig.~\ref{fig:1}), which follows the definition in \citet{shimizu2020development}. Although the implementation becomes more complex, the formulation below remains valid for more general cases involving reference planar subsurfaces or a weakly nonplanar reference surface. 
We parametrize coordinates $\boldsymbol\xi$ on $\Gamma$ as a function of the location $(x, y)$ taken on $\Gamma^\circ$. 
Specifically, $\boldsymbol\xi$ is expressed by $(x, y)$ and the elevation $z(x,y)$ of $\Gamma$ relative to $\Gamma^\circ$, where the positive $z$-direction aligns with ${\bf n}^\circ$. 

We expand the potency density using basis functions that depend only on $(x,y)$. 
The spatial distribution of the potency density tensor $D$ is represented by bases $X_j(x,y)$ ($j=1,...,M$) with expansion coefficients $a_{j}$ stored in an array ${\bf a}$: 
\begin{equation}
    D(\boldsymbol\xi,\tau)=\sum_{j=1}^M a_{j} X_j(x,y).
    \label{eq:basisfuncexp_PDT}
\end{equation}
Substituting it into eq.~(\ref{eq:RepTheorem}) yields
\begin{equation}
    u_i=H_{ij}(\Gamma)a_j,
    \label{eq:discreteObseq}
\end{equation}
where $H_{ij}(\Gamma)=\int_\Gamma d\Sigma G_i(\boldsymbol\xi)X_j(x,y)$. 

We remark that $H_{ij}(\Gamma)$ still depends on the unknown $\Gamma$ through its domain of the surface integral. In the PDTI, $H_{ij}(\Gamma)$ is commonly approximated by a known surface $\Gamma^\prime$ (e.g., $\Gamma^\circ$) or an estimate $\hat \Gamma$ [i.e., $H(\Gamma)\simeq H(\hat \Gamma)$, see \S\ref{sec:probabilisticShimizu2021} for details]. 

In this manner of eq.~(\ref{eq:discreteObseq}), the estimation target of the PDTI is decomposed into two distinct sets of model parameters: the potency coefficients ${\bf a}$ and the geometry $\Gamma$ represented by $z(\cdot,\cdot)$. 
In doing so, while the two sets of parameters remain physically interdependent, the circularity in their definitions is resolved.
The physical dependency between the two is enforced through an additional constraint, which is explored in the next section. 

\section{Formulation of inverse mapping from potency density to fault shape}
\label{sec:Gammaprior}

Since the potency density, here expressed by ${\bf a}$, determines the $n$-vector field of the earthquake fault, the essence of the reconstruction problem ${\bf a} \mapsto \Gamma$ is determining the specific function $g$ that recovers the surface shape from the $n$-vector field such that $\Gamma = g(n)$. This $n\mapsto \Gamma$ operation is the inverse to the mapping in eq.~(\ref{eq:defofnvector}) that defines an $n$-vector field for a given surface $\Gamma$ ($\Gamma\mapsto n$). 
Currently, the $n$-vector is uniquely determined for a given ${\bf a}$ by selecting the nodal plane closest to the reference vector $n^\circ$. Cases in which the $n$-vector is given probabilistically from ${\bf a}$ can be handled as an extension of this formulation.

\begin{figure}
 \includegraphics[width=90mm]{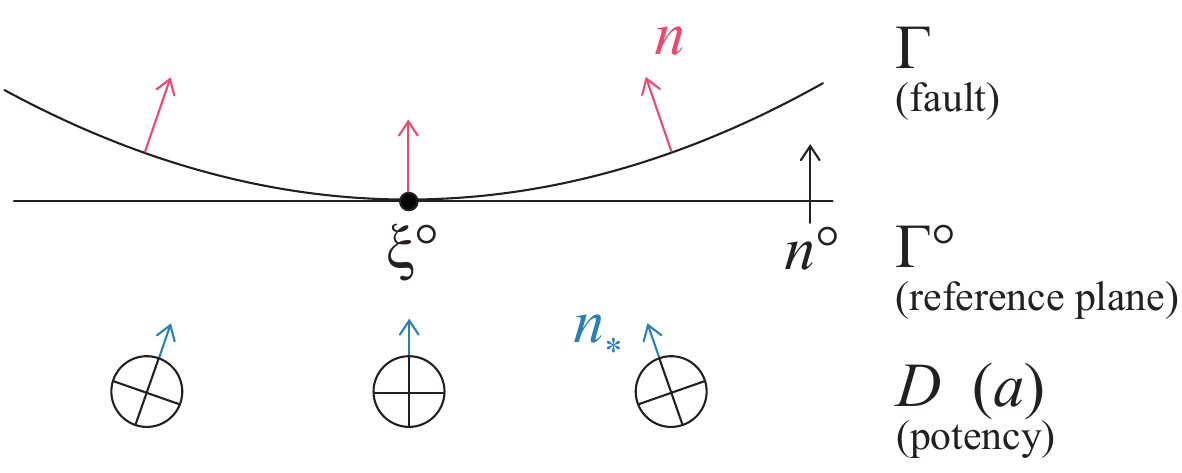}
 \caption{
 Three types of $n$-vector characterizations based on different fault properties, indicated by arrows of different colors. 
 (Red) The normal-vector field of a fault surface $\Gamma$, denoted by $n$.
 (Black) 
 The reference normal $n^\circ$, which is the $n$-vector of the reference surface $\Gamma^\circ$. 
 (Blue) The normal-vector field $n_*$ expected from the potency $D$, or precisely, from $a$ ($D$ parametrized by a basis function expansion along the reference surface.)}
 \label{fig:2}
\end{figure}

Note that the $n$-vector is doubly determined by the potency-density tensor $D$ and the fault geometry $\Gamma$ (Fig.~\ref{fig:2}). 
According to eq.~(\ref{eq:defofnvector}), the definition of the $n$-vector, the relative coordinate field $\boldsymbol\xi=(x,y,z)$ that describes $\Gamma$, namely the fault shape, uniquely sets the field of ${\bf n}$, while according to eq.~(\ref{eq:defofnvector_PDTI}), the representation of the potency density for a displacement-discontinuity surface, the tensor shape of the potency $D$ dictates ${\bf n}$. 
Those two types of $n$ are naively identical. 
When the $n$-vectors defined by $\Gamma$ and by $D$
are separately denoted by ${\bf n}(\boldsymbol\xi;\Gamma)$ 
and ${\bf n}_*(\boldsymbol\xi;{\bf a})$, respectively, this expectation is articulated as
\begin{equation}
    n(\boldsymbol\xi;\Gamma)=n_*(\boldsymbol\xi;{\bf a}).
    \label{eq:nvector_equality}
\end{equation}
Equation~(\ref{eq:nvector_equality}) is a principle in slip inversion, where $\Gamma$ is defined first and $n(\boldsymbol\xi;\Gamma)$ of $\Gamma$ constrains $n_*(\boldsymbol\xi;{\bf a})$, the tensorial shape of the potency; this principle is followed even in varying geometry~\citep[e.g.,][]{fukahata2008non,dutta2021simultaneous}. 
In the framework of the PDTI, which permits general inelastic strain as earthquake sources, the inelastic strain is expressed as the potency density $D$ first, and a nodal-plane direction $n_*(\boldsymbol\xi;{\bf a})$ of $D$ provides the information on $n(\boldsymbol\xi;\Gamma)$ via eq.~(\ref{eq:nvector_equality}). 
Appreciating these ostensibly similar approaches is fundamental to reconstructing earthquake faults from the observed seismic moment, as discussed below.

By strictly adhering to the theoretical relationship of eq.~(\ref{eq:nvector_equality}), analogous to slip inversion, 
\citet{shimizu2021construction} reconstructed $\Gamma$ from the potency ${\bf a}$ parametrized along a reference plane. 
This transformation from the $n$-vector field to the surface is facilitated by the rewritten form of eq.~(\ref{eq:nvector_equality}); 
considering the infinitesimal $dz$ in $z$ associated with the infinitesimal $d\alpha$ in $\alpha=x,y$,
\begin{equation}
    z_{,\alpha}=-n_\alpha/n_z,
    \label{eq:elevationgradientvsnormal}
\end{equation}
where the subscript following a comma denotes the partial derivative. 
The assumption of one-to-one mapping from the potency to the $n$-vector, which is equivalent to the constraint that the $n$-vector lies within 45 degrees from $n^\circ$ (\S\ref{subsec:forward}), here makes the elevation $z$ a single-valued function of $x$ and $y$.

Equation~(\ref{eq:elevationgradientvsnormal}) states that the cotangent of the $n$-vector corresponds to the downward gradient of a curve in a two-dimensional space, where the slip direction perpendicular to the $n$-vector is exactly the fault orientation. 
Accordingly, \citet{shimizu2021construction} averaged the potency along the short axis of the reference surface and smoothly connected the resulting $n$-vector along the long axis to extract the two-dimensional features of a uniaxially bent fault (e.g., along-strike strike angle variations and along-dip dip angle variations). 
The same procedure adhering to eq.~(\ref{eq:nvector_equality}) can also reproduce quasi-two-dimensional shapes including uniaxial twists such as along-strike dip variations and along-dip strike variations (\S\ref{subsec:synthetic3}). This study regards such surface reconstruction fully governed by eq.~(\ref{eq:nvector_equality}), including quasi-two-dimensional surface reconstruction, as falling within the scope of the previous method by \citet{shimizu2021construction}. 

Crucially, this strategy adhering to eq.~(\ref{eq:nvector_equality}) does not extend to three-dimensional surface reconstruction, a limitation attributed to the difference in the degrees of freedom of $n$-vectors between two- and three-dimensional spaces. 
In two dimensions, the $n$-vector is parametrized by the polar angle $\theta$. 
The transformation of a given $n$-vector field to an elevation $z$ field is then a one-to-one mapping $\theta\mapsto z$, which provides a unique $z$ field to an arbitrary smooth $n$ field. 
By contrast, in three dimensions, 
the $n$-vector may be expressed by the polar angle $\theta$ and the azimuthal angle $\phi$. 
Three-dimensional surface reconstruction from the $n$-vector is therefore a two-to-one mapping $(\theta,\phi)\mapsto z$ from a vector field of $\theta$ and $\phi$ to a scalar field of $z$, generally having no solutions for eqs.~(\ref{eq:nvector_equality}) and (\ref{eq:elevationgradientvsnormal}) unless the given $n$-vector field is error-free. 
Namely, the surface reconstruction from potency is an overdetermined problem in three dimensions. This overdeterminacy derives from the definition of the $n$-vector eq.~(\ref{eq:defofnvector}).

This overdeterminacy can be intuitively grasped from the perspective of the one-to-two mapping eq.~(\ref{eq:elevationgradientvsnormal}) from $z$ to $(\theta,\phi)$, which here precisely corresponds to the definition of the normal vector eq.~(\ref{eq:defofnvector}), made bijective by the smoothness of the surface. 
A smooth surface with no tears, the aforementioned regular surface, sets any closed-path integral of relative elevation on a reference surface to zero: 
\begin{equation}
    \oint d{\bf l}\cdot \nabla z=0,
\end{equation}
where $\nabla=(\partial_x,\partial_y)$ denotes the two-dimensional partial differential operator. 
Stokes' theorem rewrites it as
\begin{equation}
    \int dxdy \nabla\times \nabla z=0,
\end{equation}
that is,
\begin{equation}
    \partial_x\partial_y z=\partial_y\partial_x z.
    \label{eq:commutability}
\end{equation}
Under this constraint of eq.~(\ref{eq:commutability}) on $z$, the one-to-two mapping from $z$ to $(\theta,\phi)$ becomes bijective.
Equation~(\ref{eq:commutability}) indicates that 
the condition for the smooth surface free of tears is the integrability of $\nabla z$, that is, the elevation $z$ of a tear-free surface is a continuous function with no jumps.
Quasi-two-dimensional surface reconstruction is an approximation that neglects that the slope along the long axis may vary along the short axis (e.g. $\partial_y z_{,x}\simeq 0$), 
so it violates eq.~(\ref{eq:commutability}) when the slope along the short axis varies along the long axis (e.g. $\partial_x z_{,y}\neq 0$). 

Substituting eq.~(\ref{eq:elevationgradientvsnormal}) into eq.~(\ref{eq:commutability}) yields 
\begin{equation}
    \partial_x(n_y/n_z)=
    \partial_y(n_x/n_z).
    \label{eq:commutability_constraint_nvect}
\end{equation}
Two-dimensional shapes of uniaxial bending called single-curved surfaces, treated in \citet{shimizu2021construction}, always set both sides of eq.~(\ref{eq:commutability_constraint_nvect}) to zero. 
However, only the error-free $n$-vector field of a surface meets eq.~(\ref{eq:commutability_constraint_nvect}) in three dimensions. 
Other than such special cases, the estimated normal vectors of nonplanar surfaces violate the integrability condition of eq.~(\ref{eq:commutability_constraint_nvect}), and that is the reason why the $(\theta,\phi)\mapsto z$ mapping of eqs.~(\ref{eq:nvector_equality}) and (\ref{eq:elevationgradientvsnormal}) is generally overdetermined: 
\begin{equation}
    \partial_x(n_{*y}/n_{*z})\neq
    \partial_y(n_{*x}/n_{*z}).
    \label{eq:noncommutabilitynstar}
\end{equation}
An $n$-vector field randomly perturbed at individual coordinates does not satisfy the non-local gradient constraint eq.~(\ref{eq:noncommutabilitynstar}) between adjacent points, resulting in pervasive tearing of the reconstructed surface.

Since the source surface reconstruction via eqs.~(\ref{eq:nvector_equality}) and (\ref{eq:elevationgradientvsnormal}) is overdetermined and eq.~(\ref{eq:elevationgradientvsnormal}) is the identity of the normal vector of a surface, surface reconstruction from given potency must read
eq.~(\ref{eq:nvector_equality}) as some projection:
\begin{equation}
    n(\boldsymbol\xi;\Gamma)\xrightarrow{\mathcal P} n_*(\boldsymbol\xi;{\bf a}),
    \label{eq:nvector_limitequality}
\end{equation}
where $\xrightarrow{\mathcal P}$ denotes some operation to remove non-integrable components such that $\partial_x(n_{*y}/n_{*z})\neq \partial_y(n_{*x}/n_{*z})$. 
Only if that non-integrable portion of $n_*$ is removed, that is, only if illogical tears are ruled out, 
is a fault reconstructed as a smooth surface in a coarse-grained continuum.

Because $n\mapsto z$ mapping is impossible when the theoretical relationship between potency and shape of eq.~(\ref{eq:nvector_equality}) is required to strictly hold, 
surface reconstruction from potency must relax this premise. There remains, however, arbitrariness in eq.~(\ref{eq:nvector_limitequality}) on how that relaxation is applied. We describe this subjectively introduced criterion as prior information in the sense of Bayes. 
Equation~(\ref{eq:nvector_limitequality}) expects that the $n$-vector ${\bf n}(\boldsymbol\xi,\Gamma)$ of a surface $\Gamma$ approaches the $n$-vector ${\bf n}_*({\bf a})$ (defined by eq.~\ref{eq:defofnvector_PDTI}) pointed by the nodal planes of potency beachballs. 
The constraint of approximating a normalized vector to a designated value yields a non-Gaussian central-limit distribution known as the von Mises-Fisher distribution; it measures the distance of vectors by the inner product and takes the following form for a uniform isotropic case: 
\begin{equation}
    P(\Gamma|{\bf a};\kappa)=\exp[\kappa\int_\Gamma d\Sigma(\boldsymbol\xi) {\bf n}_*(\boldsymbol\xi,{\bf a})\cdot {\bf n}(\boldsymbol\xi,\Gamma)]/\mathcal Z,
    \label{eq:prior_Gamma}
\end{equation}
where $\kappa$ denotes a scale factor, and $\mathcal Z$ is a normalization constant. 
The domain of integration is set to $\Gamma$, such that $P(\Gamma|{\bf a};\kappa)$ is independent of the way to take the reference surface $\Gamma^\circ$. 

Then, eq.~(\ref{eq:nvector_limitequality}) requires the limit that places ${\bf n}$ in the integrable nearest neighbor of non-integrable ${\bf n}_*$, 
that is, the displacement discontinuity that best approximates given inelastic strain. 
It is termed dislocation limit hereafter. In eq.~(\ref{eq:prior_Gamma}), the dislocation limit is described as
\begin{equation}
    \kappa\to\infty.
\end{equation}
In this limit,
\begin{equation}
    P(\Gamma|{\bf a})=\delta(\Gamma-\Gamma_*({\bf a})),
    \label{eq:prior_Gamma_limit}
\end{equation}
where $\delta(\cdot)$ denotes Dirac delta and 
\begin{equation}
    \Gamma_*:={\rm argmax}_{\Gamma|{\bf a}}P(\Gamma|{\bf a};\kappa).
    \label{eq:defofGammastar}
\end{equation}
Here ${\rm argmax}$ denotes the arguments of the maxima. 
In this paper, $P(\Gamma|{\bf a})$ is taken as $\lim_{\kappa\to\infty}P(\Gamma|{\bf a},\kappa)$.

When the non-integrable part $(\partial_x\partial_y -\partial_y\partial_x )z$ expected from $n_*$ is small, the vector $n$ obtained from eq.~(\ref{eq:nvector_limitequality}) does not deviate significantly from $n_*$. 
Basically, we suppose this regime and proceed by adopting the prior $P(\Gamma|{\bf a})$ written as eq.~(\ref{eq:prior_Gamma}). 
However, the overconstraint in surface reconstruction, where no smooth surface satisfies the forward model eq.~(\ref{eq:defofnvector_PDTI}) that leads to eq.~(\ref{eq:nvector_equality}), raises the question of what priors $P(\Gamma|{\bf a})$ are appropriate
to relax eq.~(\ref{eq:defofnvector_PDTI}) when the non-integrability of $n_*$ is not small. 
It is important to note that the a priori constraint here is imposed on $\Gamma$ given ${\bf a}$, rather than on the inelastic strain ${\bf a}$ distributed over a given surface $\Gamma$ as in slip inversion. Considering that ${\bf a}$ represents the $n$-vector, this approach of $\Gamma|{\bf a}$ implies a reversal of the standard definition of the $n$-vector, where $n$ is defined for $\Gamma (z)$. The PDTI is commonly formulated in such a way that determines $\Gamma$ from ${\bf a}$; ${\bf a}$ is inferred by using another prior constraint independent of $\Gamma$. $P(\Gamma|{\bf a})$ introduced here was designed to align with this structure. 
While this approach reverses the standard definition of the $n$-vector, the physical nature of faulting supports this causal order of $\Gamma|{\bf a}$ rather than contradicts it.
Even before the $n$-vector is disturbed by measurement errors that do not satisfy eq.~(\ref{eq:nvector_equality}), 
an earthquake source as a surface is an approximation of a fracture zone, an idealization following the nature of brittle failure localized within a quasi-face~\citep{matsu2019physical}, 
so the exact observance of eq.~(\ref{eq:defofnvector_PDTI}) is unphysical from the outset. 
Thus, if read as a forward model that precedes the observation, $P(\Gamma|{\bf a})$ that determines a surface from a given $n$-vector field is a representation of the ambiguity in defining an earthquake fault surface. This surface is an approximation for a thin sheet of inelastic deformation. 

In summary, the potency density that best explains the data retains information that cannot be mapped back onto a displacement discontinuity, a limit of inelastic strain strictly localized on a surface. The resulting surface is forced to violate the integrability of the fault slope, which means the non-integrability of the potency density, the dislocation. 
Briefly, the best earthquake source is generally not a displacement discontinuity.
In conventional point-source representations, the discrepancy between a six-degree-of-freedom moment tensor and a five-degree-of-freedom dislocation is resolved locally by simply discarding the eigenvalue with the smallest absolute magnitude. By contrast, in an areal finite-source representation, this dimensionality gap manifests as a non-local geometric conflict. Physically, when a normal vector $n$ is assigned to each coordinate based on local potency, the resulting field fails to satisfy the gradient-based constraints that couple adjacent points. The reconstructed source thus represents not a smooth, continuous fault, but a collection of infinitesimal fragments that are ``torn everywhere'', a fragmented geometry where the dislocation is nowhere integrable (eq.~\ref{eq:noncommutabilitynstar}). Therefore, reconstructing a meaningful fault geometry requires more than a local approximation, necessitating a global approach to the overdeterminacy inherent in the transition from an areal inelastic strain field to a displacement-discontinuity surface. 
Equation (\ref{eq:commutability_constraint_nvect}) represents the integrability condition necessary to project the areal inelastic strain onto displacement discontinuity without loss of information.
It ensures the global consistency of the normal vector field, functioning as the areal-source counterpart to the determinant-free condition in point-source analysis.

Therefore, we must distinguish between two types of smoothness regarding fault geometry: the smoothness of the surface gradient (i.e., the potency density field) and the smoothness of the surface itself. We have revealed that these are not equivalent. Even if the normal vector field varies smoothly, the corresponding elevation derivatives do not necessarily satisfy the integrability condition; thus, a smooth normal vector field generally fails to form a smooth, continuous surface (eq.~\ref{eq:noncommutabilitynstar}). We have formulated the criterion embodying the assumption that a fault constitutes a smooth surface (eq.~\ref{eq:prior_Gamma}). 
Consequently, relaxing the smoothness assumption involves two distinct levels: a `weak' assumption regarding the smoothness of the normal vector, and a `strong' assumption regarding the smoothness of the surface continuity. Our approach to inferring macroscopic fault geometry from limited data implicitly accepts the weak assumption of potency smoothness, the validity of which lies within the domain of potency estimation. The mathematics of the $n$-vector-geometry mapping---specifically, that the surface `tears' even presenting a smooth potency field---proves that a smooth potency-density field can convey information on macroscopic discontinuities, tears of faults, even under the weak smoothness assumption. After all, $n$-vector components violating the slope integrability condition (eq.~\ref{eq:commutability_constraint_nvect}) are not necessarily illogical artifacts but may represent valid information regarding fault geometry discontinuity. Surface reconstruction explicitly incorporating such fault discontinuities is left for future study.

Note that $P(\Gamma|{\bf a})$ is different from the posterior probability of fault shape. The mapping from the $n$-vector to geometry, and consequently the surface reconstruction of earthquake faults from potency, never transcends the subjective description of Bayesian probability even when potency is objectively known.
The posterior of fault shape is the convolution between the conditional prior of fault shape given potency $P(\Gamma|{\bf a})$ and the posterior of potency, and the dislocation limit reduces the posterior of the PDTI to that in slip inversion (Appendix \ref{app:A}). 


\section{Derivation of the optimal elevation field $\Gamma_*$}

We have introduced $P (\Gamma|{\bf a})$ by recognizing the subjective nature of the mapping from two degrees of freedom in the expected normal $n_*$ to one degree of freedom in the smooth surface elevation. 
With the prior distribution $P (\Gamma|{\bf a})$ exemplified by eq.~(\ref{eq:prior_Gamma}), we derive an analytical expression of the most probable surface estimate $\Gamma_*({\bf a})$ given the potency ${\bf a}$. 
In the previous section, $P(\Gamma|{\bf a})$ was constructed exclusively based on the $n$-vector identity. This formulation is equivalent to a free boundary condition, where the partial differential equation is solved without external boundary constraints. We mainly analyze this case. A generalization to cases with imposed boundary conditions will be addressed at the end of this section.

Since we assumed the case where the potency density ${\bf a}$ uniquely determines the field of the normal vector $n_*$ (\S\ref{subsec:forward}), the goal of this section is to obtain $\Gamma_*$ given $n_*$. 
The $xy$-plane here represents the coordinate space spanned along the reference surface $\Gamma^\circ$, where $\Gamma$ is parametrized by the elevation $z(\cdot,\cdot)$ of $\Gamma$ relative to $\Gamma^\circ$. 
In this section, we assume that the non-integrable component of $n_*$ (eq.~\ref{eq:noncommutabilitynstar}) is sufficiently small so that the solution of surface reconstruction is insensitive to the functional form of $P (\Gamma|{\bf a})$. The full-order analysis appears in Appendix~\ref{app:B}.

\subsection{Governing equation}
\begin{figure}
 \includegraphics[width=90mm]{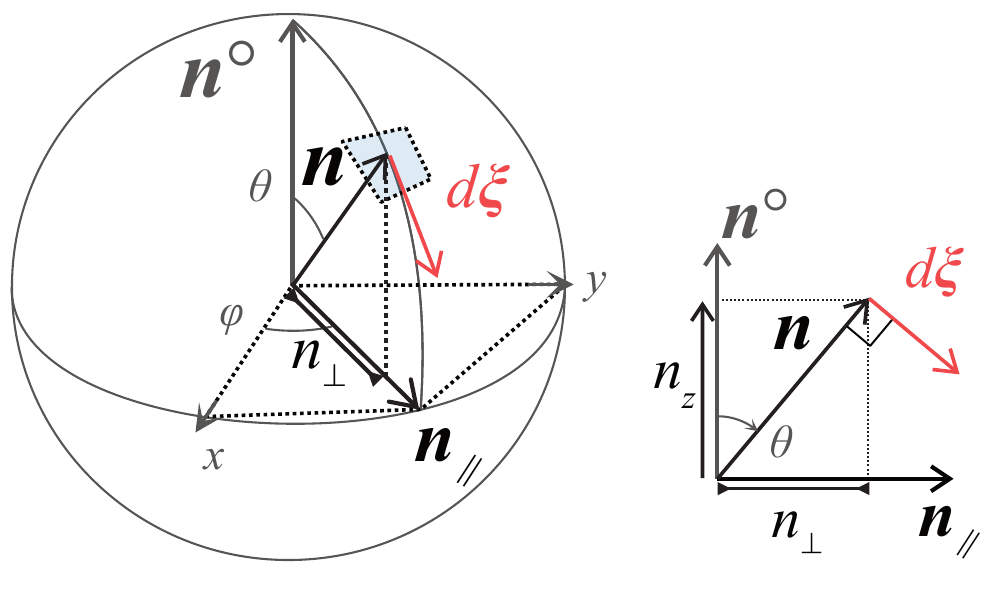} 
 \caption{
 Geometric relationship between the normal vector ${\bf n}$ and the tangent plane $d\boldsymbol\xi$ of a surface. 
 A Cartesian coordinate is defined such that the reference-surface normal ${\bf n}^\circ$ aligns with the $z$ axis. The $xy$-plane represents the coordinate space spanned along the reference surface. 
 The horizontal projection ${\bf n}_\parallel$ of ${\bf n}$ indicates the surface slope direction, while the ratio of the horizontal component $n_\perp$ to the vertical component $n_z$ of ${\bf n}$ determines the downward surface slope. 
 }
 \label{fig:4}
\end{figure}

The governing equation for the problem is the definitional identity of the $n$-vector (eq.~\ref{eq:defofnvector}, simplified to eq.~\ref{eq:elevationgradientvsnormal}), 
which expresses the orthogonality between the $n$-vector and the nonplanar surface. 
Defining
\begin{flalign}
    n_\perp&=\sqrt{n_x^2+n_y^2}
    \\
    {\bf n}_\parallel&=(n_x,n_y)/n_\perp,
\end{flalign}
we can rewrite eq.~(\ref{eq:elevationgradientvsnormal}) as
\begin{equation}
    \nabla z = - \frac{n_\perp}{n_z}{\bf n}_\parallel.
\label{eq:elevationgradientvsnormal_meaning}
\end{equation}
As shown in eq.~(\ref{eq:elevationgradientvsnormal_meaning}), eq.~(\ref{eq:elevationgradientvsnormal}) represents more than that a 90-degree rotation (cotangent, $n_\perp/n_z$) of the $n$-vector gives the surface slope implied in two dimensions. 
The $n$-vector indicates the direction of the slope by its horizontal projection ${\bf n}_\parallel$ ($|{\bf n}_\parallel|=1$) while quantifying the downward slope by its horizontal-vertical ratio $n_\perp/n_z$ (Fig.~\ref{fig:4}).

The subsequent analysis is simplified by parametrizing the $n$-vector using hemispherical coordinates. Without loss of generality, the angle between ${\bf n}$ and ${\bf n}^\circ$ is supposed to be within $[0,\pi/2]$. 
Let the polar angle $\theta\in[0,\pi/2]$ be the angle between ${\bf n}$ and ${\bf n}^\circ$, and 
the azimuthal angle $\phi\in[0,2\pi)$ be measured from the $x$ axis. In this setting, the components of ${\bf n}$ are expressed as
\begin{equation}
    \begin{aligned}
        n_z&=\cos\theta
        \\
        n_\perp&=\sin\theta
        \\
        n_{\parallel x}&=\cos\phi
        \\
        n_{\parallel y}&=\sin\phi.
    \end{aligned}
    \label{eq:ns_onsurface}
\end{equation}
Note $n_x=\sin\theta\cos\phi$ and $n_y=\sin\theta\sin\phi$. 
Equations~(\ref{eq:ns_onsurface}) reduce eq.~(\ref{eq:elevationgradientvsnormal_meaning}) to 
\begin{equation}
    \nabla z=-\tan\theta
    \left(
    \begin{array}{c}
    \cos\phi
    \\
    \sin\phi
    \end{array}
    \right).
    \label{eq:elevationgradientvsnormal_spherical}
\end{equation}
Hereafter, the local basis vectors associated with $(\theta,\phi)$ on the unit hemisphere associated with ${\bf n}$ are denoted as ${\bf n}$ (radial), ${\bf e}_{\theta}$ (polar), and ${\bf e}_{\phi}$ (azimuthal). 

\subsection{Solving a stochastic partial differential equation for fault elevation}
\label{subsec:solvingSPDE_elevation}

The geometry $\Gamma$ of a smooth surface is equivalent to both the elevation field $z$ and the normal-vector field $n$, but formally, eq.~(\ref{eq:elevationgradientvsnormal}) can be read as a partial differential equation of $z$ with a random external force $-(n_\perp/n_z){\bf n}_\parallel$ imposed. 
The probability distribution $P(\Gamma|{\bf a};\kappa)$ (eq.~\ref{eq:prior_Gamma}) describes the fluctuations of the $n$-vector around its expectation $n_*$. 
Here, we derive an analytical solution of this stochastic partial differential equation of eqs.~(\ref{eq:elevationgradientvsnormal}) and (\ref{eq:prior_Gamma}). Strictly speaking, the presence of significant non-integrable components (eq.~\ref{eq:noncommutabilitynstar}) of $n_*$ necessitates an explicit operation to remove them. However, since the results of this operation depend on the details of the prior $P(\Gamma|{\bf a};\kappa)$, we consider such scenarios out of scope for the present analysis to maintain generality. We focus on the case where $n \simeq n_*$ holds such that the functional form of the prior is irrelevant. The full-order derivation addressing the higher-order complexity is provided in Appendix \ref{app:B}.

Although eq.~(\ref{eq:prior_Gamma}) is a central-limit distribution, it is non-Gaussian due to the $n$-vector normalization, not solvable via linear operations. 
The explicit form of this distribution is obtained by writing the $n$-vector as a function of $\nabla z$ (eq.~\ref{eq:nvectorfromslopes} in the next section) in eq.~(\ref{eq:prior_Gamma}). 
However, in the dislocation limit, those non-Gaussian fluctuations become infinitesimal, and the distribution reduces to a Gaussian peaked at $\Gamma_*$, which we evaluate below.

First, we rewrite the Cartesian forms of $P(\Gamma|{\bf a};\kappa)$ (eq.~\ref{eq:prior_Gamma}) in hemispherical coordinates. 
In a hemispherical coordinate system at a unit hemispherical position ($\theta_*$, $\phi_*$) of ${\bf n}_*$, ${\bf n}$ is expressed as 
\begin{equation}
    {\bf n}=\chi_n^*{\bf n}_*+\chi_\theta^*{\bf e}_{\theta}^*+\chi_\phi^*{\bf e}_{\phi}^*,
    \label{eq:n_series_bystars}
\end{equation}
with associated coefficients ($\chi_n^*$, $\chi_\theta^*$, $\chi_\phi^*$). 
Since $\partial {\bf n}(\theta,\phi)/\partial \theta={\bf e}_\theta$ and $\partial {\bf n}(\theta,\phi)/\partial \phi={\bf e}_\phi\sin\theta$, the first order of the Taylor series with respect to angle fluctuations $\theta-\theta_*$ and $\phi-\phi_*$ around $n_*$ yields 
\begin{equation}
    \begin{aligned}        
    \chi_\theta^*&=(\theta-\theta_*)+\mathcal O(\delta\Omega^2)
    \\
    \chi_\phi^*&=(\phi-\phi_*)\sin\theta_*+\mathcal O(\delta\Omega^2),
    \end{aligned}
\end{equation}
where
$\mathcal O(\delta \Omega^k)$ denotes the $k$-th order angle fluctuations from $n_*$. 
From the normalization condition ${\bf n}\cdot{\bf n}=1$ of 
${\bf n}$ and the above expressions of $\chi_\theta^*$ and $\chi_\phi^*$, the remaining coefficient $\chi_n^*$ is evaluated as 
\begin{equation}
    \chi_n^*=1-[(\theta-\theta_*)^2+(\phi-\phi_*)^2\sin^2\theta_*]/2+\mathcal O(\delta\Omega^3).
    \label{eq:n_cdot_nstar}
\end{equation}
Substituting eq.~(\ref{eq:n_cdot_nstar}) into the von Mises-Fisher distribution of eq.~(\ref{eq:prior_Gamma}), we obtain
\begin{equation}
    {\bf n}_*\cdot {\bf n}=1-[(\theta-\theta_*)^2+(\phi-\phi_*)^2\sin^2 \theta_*]/2+\mathcal O(\delta\Omega^3)
    \label{eq:nnstarinnerproduct}
\end{equation}
and 
\begin{equation}
    P(\Gamma|{\bf a};\kappa)=\exp\{
    -\frac\kappa2 \int_\Gamma d\Sigma[ (\theta-\theta_*)^2+(\phi-\phi_*)^2\sin^2\theta_*+\mathcal O(\delta\Omega^3)]\}/\mathcal Z^\prime,
    \label{eq:Quadraticsolidangle}
\end{equation}
where $\mathcal Z^\prime$ represents normalization. The lowest order of $P(\Gamma|{\bf a};\kappa)$ is a simple Gaussian, and higher-order terms are negligible in the dislocation limit $\kappa\to\infty$, as long as non-integrable components (eq.~\ref{eq:noncommutabilitynstar}) of $n_*$ are small (Appendix \ref{app:B}). 

Next, we evaluate the partial differential equation for the elevation (eq.~\ref{eq:elevationgradientvsnormal}), expressed in hemispherical coordinates as eq.~(\ref{eq:elevationgradientvsnormal_spherical}). 
Linearizing eq.~(\ref{eq:elevationgradientvsnormal_spherical}) with respect to angle fluctuations around $n_*$ gives
\begin{equation}
    \nabla z=\nabla z_*-\cos^{-2}\theta_*
    \left(
    \begin{array}{c}
    \cos\phi_*
    \\
    \sin\phi_*
    \end{array}
    \right) (\theta-\theta_*)-\tan\theta_*    
    \left(
    \begin{array}{c}
    -\sin\phi_*
    \\
    \cos\phi_*
    \end{array}
    \right) (\phi-\phi_*)+\mathcal O(\delta\Omega^2),
\end{equation}
where $z_{*,\alpha}:=-n_\alpha/n_z$. 
Projecting the difference $\nabla z-{\nabla z}_*$ onto ${\bf n}_{*\parallel}$ and its orthogonal direction using eqs.~(\ref{eq:ns_onsurface}), we find
\begin{flalign}
    \theta-\theta_* &=-n_{*z}^2{\bf n}_{*\parallel} \cdot (\nabla z-{\nabla z}_*)+\mathcal O(\delta\Omega^2)
    \\
    \phi-\phi_*&= -(n_{*z}/ n_{*\perp}){\bf n}_{*\parallel}\times (\nabla z-{\nabla z}_*)+\mathcal O(\delta\Omega^2), 
\end{flalign}
or explicitly, recalling ${\bf n}_\parallel\cdot \nabla z=-n_\perp/n_z$ and ${\bf n}_\parallel\times \nabla z=0$ (eq.~\ref{eq:elevationgradientvsnormal_meaning}),
\begin{flalign}
    \theta-\theta_*&= -n_{*z}^2( n_{*\parallel x} z_{,x}+n_{*\parallel y} z_{,y}+n_{*\perp}/n_{*z})+\mathcal O(\delta\Omega^2)
    \label{eq:deltatheta}
    \\
    \phi-\phi_*&= -n_{*\perp}^{-1} n_{*z}(n_{*\parallel x} z_{,y}- n_{*\parallel y} z_{,x})+\mathcal O(\delta\Omega^2).    
    \label{eq:deltaphi}
\end{flalign}

Plugging eqs.~(\ref{eq:deltatheta}) and (\ref{eq:deltaphi}) into 
the exponent of $P(\Gamma|{\bf a})$ (eq.~\ref{eq:Quadraticsolidangle} with $\kappa\to\infty$), 
we arrive at
\begin{equation}
\begin{aligned}
    \ln [P(\Gamma|{\bf a})\times \mathcal Z^\prime]=
- \frac\kappa 2 \left\{\int_\Gamma d\Sigma 
[
n_{*z}^2( n_{*\parallel x} z_{,x}+n_{*\parallel y} z_{,y}+n_{*\perp}/n_{*z})]^2
\right.
\\
\left.
+
\int_\Gamma d\Sigma 
[n_{*z}(n_{*\parallel x} z_{,y}- n_{*\parallel y} z_{,x})
]^2+\mathcal O(\delta\Omega^3)]
\right\}.
\end{aligned}
\label{eq:Gammaprior_lowestorder}
\end{equation}
Here $\sin\theta=n_\perp$ (in eqs.~\ref{eq:ns_onsurface}) is also used.
Note that the area element $d\Sigma(\boldsymbol\xi)$ along the fault is larger than the area element $dxdy$ along the reference plane because of the tilt of the fault normal by $\theta$: 
\begin{equation}
d\Sigma(\boldsymbol\xi)=\frac{dxdy}{|\cos\theta(\boldsymbol\xi)|}=|n_z|^{-1}dxdy.
\label{eq:areaelementconversion}
\end{equation}
We remark that $|n_z|=n_z$ in the hemispherical coordinates. 
In the dislocation limit, with small non-integrable components (eq.~\ref{eq:noncommutabilitynstar}) of $n_*$, the higher order fluctuations from $n_*$ are dropped, and the most probable estimate $\Gamma_*$ for given $n_*$ is finally expressed as:
\begin{equation}
\begin{aligned}    
    \Gamma_*={\rm argmin}\left\{\int   
|n_{*z}|( n_{*\parallel x}n_{*z} z_{,x}+n_{*\parallel y}n_{*z} z_{,y}+n_{*\perp})^2dxdy
\right.\\\left.
+
\int 
|n_{*z}|(n_{*\parallel x} z_{,y}- n_{*\parallel y} z_{,x})
^2dxdy
\right\},
\end{aligned}
\label{eq:optimaloptimal_explicit}
\end{equation}
where ${\rm argmin}$ denotes the arguments of the minima. 
To avoid zero divisions at $n_{*z}=0$, 
$n_{*z}^2$ among the weight $|n_{*z}^3|$ in the first term is absorbed into the brackets. 
The first term from the $\theta$ fluctuation constrains the fault slope $\nabla z$ component parallel to ${\bf n}_\parallel$, while the second term from the $\phi$ fluctuation constrains the $\nabla z$ component perpendicular to ${\bf n}_\parallel$. 
Their weights $|n_{*z}^3|$ and $|n_{*z}|$, which derive from 
the rotational invariance of $P(\Gamma|{\bf a})$ with respect to the reference plane, weaken the constraints for steeper slopes (i.e., smaller $n_{*z}$ for larger $\theta(<\pi/4)$: $\partial n_{*z}/\partial\theta_*=-\sin\theta_*<0$).

The problem posed by eqs.~(\ref{eq:elevationgradientvsnormal})
and (\ref{eq:prior_Gamma}) is thus reduced to the least-square minimization in eq.~(\ref{eq:optimaloptimal_explicit}). 
It is worth noting that the minimization function in eq.~(\ref{eq:optimaloptimal_explicit}) depends only on the derivatives of $z$, which leaves the absolute vertical position unconstrained. 
We have assumed that the absolute coordinate of one point on a surface $\boldsymbol\xi^\circ$ to determine the absolute position
(\S\ref{subsec:forward}). 
The estimated relative positions that parametrize the surface shape remain invariant regardless of how the unconstrained absolute vertical level is fixed. 

The rest of analysis is the ordinary linear inversion, which we omit to detail.
In the subsequent analysis, we have solved it using a basis function expansion with trapezoidal integration of eq.~(\ref{eq:optimaloptimal_explicit}). We remark that solving eq.~(\ref{eq:optimaloptimal_explicit}) in the continuous space using integration by parts results in a second-order partial differential equation of $z$ that depends on the $x$ and $y$ derivatives of the $n$-vector; it shows that $\Gamma_*$ follows an elliptic partial differential equation, which intrinsically prefers a smooth solution.

We have thus far considered the surface reconstruction without imposing any constraints on the surface edge, corresponding to the solution of the partial differential equation of eq.~(\ref{eq:elevationgradientvsnormal}) for the free boundary condition. Other boundary conditions, for example, to account for fault traces and to join multiple subfaults, can be treated as extensions of the above problem with additional boundary constraints. Functional forms of $P(\Gamma|{\bf a};\kappa)$ may be modified in such variants. 
All those extensions are similarly solvable in the following form; by adding the loss term with respect to the absolute elevation $z$ to the orientation loss with respect to $(\theta,\phi)$ as functions of $\nabla z$ (eqs.~\ref{eq:deltatheta} and \ref{eq:deltaphi}), we can generalize $P(\Gamma|{\bf a})$ as
\begin{equation}
Q(\Gamma|{\bf a})=\exp\left[
    -\frac 1 2 \int d\Sigma(\boldsymbol\xi)\int d\Sigma(\boldsymbol\xi^\prime)
    \left(
    \begin{array}{cc}
    \theta(\boldsymbol\xi)-\theta_*(\boldsymbol\xi)
    \\
    \phi(\boldsymbol\xi)-\phi_*(\boldsymbol\xi)
    \\
    z(\boldsymbol\xi)-\bar z(\boldsymbol\xi)
    \end{array}\right)^{\rm T}
    {\bf w}(\boldsymbol\xi,\boldsymbol\xi^\prime)
        \left(
    \begin{array}{cc}
    \theta(\boldsymbol\xi^\prime)-\theta_*(\boldsymbol\xi^\prime)
    \\
    \phi(\boldsymbol\xi^\prime)-\phi_*(\boldsymbol\xi^\prime)
    \\
    z(\boldsymbol\xi^\prime)-\bar z(\boldsymbol\xi^\prime)
    \end{array}\right)
    \right]/\mathcal Z^{\prime\prime}.
    \label{eq:Quadraticsolidangle_gen}
\end{equation}
Here, $\bar z$ denotes the expected elevation field, ${\bf w}(\cdot,\cdot)$ is the inverse covariance, and $\mathcal Z^{\prime\prime}$ represents normalization. 
Note that boundary conditions exceeding the determination of the vertical translation introduce further overdeterminacy, whereby the solution varies according to the assigned weights of boundary penalties; this is because the solution of the governing equation (eq.~\ref{eq:elevationgradientvsnormal_meaning}), $z(x,y) = z(0,0) - \int_{(0,0)}^{(x,y)} d{\bf l} \cdot \frac{n_\perp}{n_z}{\bf n}_\parallel$, is unique up to the vertical offset $z(0,0)$.
The quadratic form in eq.~(\ref{eq:Quadraticsolidangle_gen}) includes forms of $P(\Gamma|{\bf a})$ other than eq.~(\ref{eq:prior_Gamma}) when the higher orders are dropped in the dislocation limit.
This generalized probability density function eq.~(\ref{eq:Quadraticsolidangle_gen}) is applicable also to surface reconstruction from the point clouds of microseismicity focal mechanisms.
In our teleseismic application, we use eq.~(\ref{eq:Quadraticsolidangle_gen}) for computational acceleration (\S\ref{sec:recipe}).

\section{Surface reconstruction from synthetic noisy normal vectors}
\label{sec:synthetic}

We conduct synthetic tests to verify the most probable surface estimation (eq.~\ref{eq:optimaloptimal_explicit}) given the normal vector field. 
Simultaneous inference of potency density and geometry often yields results similar to fault geometry reconstructed from the PDTI with fixed potency locations~\citep[also see \S\ref{eq:apptoEQ}]{shimizu2021construction}. 
Therefore, this verification of surface reconstruction from a given noisy $n$-vector field serves as a primary verification of the PDTI involving fault shape inference performed in \S\ref{sec:app}.

We first set a ground-truth elevation $z_0(x,y)$ perpendicular to the reference plane $x$-$y$ and calculate its normal vector field $n_0(x,y)$. We add a noise $\delta n(x,y)$ to $n_0(x,y)$ to synthesize noisy $n_*(x,y)$, from which we compute an estimate of the elevation $\Gamma_*$ and assess the accuracy compared to the correct shape. 
In our synthetic tests, we disturb all components of the $n$-vector by Gaussian random numbers of mean 0 and standard deviation $\sigma_n$, normalizing the resulting vectors as $n_*$. 
For simplicity, we collocate the $n_*$ values on the grid points, assuming that the grid interval is large enough to neglect spatial correlations. The grid is the same as the knot configuration in the basis function expansion of the PDTI appearing in our application (\S\ref{eq:apptoEQ}), which consists of 40 points along strike $x$ and 4 points along dip $y$ with a spacing of 5 km, totaling 160 points. There are 320 independent components of the input $n$-vectors. The length scale in the following tests is nondimensionalized by setting the unit length to 1 km. 
The elevation field is expressed by the quadratic B-spline functions, with knot intervals 1.5 times wider than the $n_*$ collocation interval.

The transformation of the fault slope to the fault normal in Cartesian coordinates is given by eq.~(\ref{eq:elevationgradientvsnormal}) and the normalization condition $|{\bf n}|^2=1$: 
\begin{equation}
    \begin{aligned}
        n_z=1/\sqrt{1+z_{,x}^2+z_{,y}^2}
        \\
        n_\alpha=-z_{,\alpha}/\sqrt{1+z_{,x}^2+z_{,y}^2}.
    \end{aligned}
    \label{eq:nvectorfromslopes}
\end{equation}
For hemispherical coordinates, 
from eq.~(\ref{eq:elevationgradientvsnormal_spherical}), 
\begin{equation}
    \begin{aligned}
        \theta={\rm atan}\sqrt{z_{,x}^2+z_{,y}^2}
        \\
        \phi={\rm atan}2(z_{,y},z_{,x}),
    \end{aligned}
\end{equation}
where ${\rm atan}2(\cdot,\cdot)$ denotes 2-argument arctangent.

We treat two model shapes.
The first shape is a uniaxial twist (\S\ref{subsec:synthetic1}). 
This shape is characterized by biaxial slopes, providing a simplest instance of a torn surface (i.e. eq.~\ref{eq:commutability_constraint_nvect} violation) if a uniaxial interpolation is applied to the $n$-vector field. 
The uniaxial twist also represents a developable surface, the surface that can be exactly expressed by spline interpolation without error. These make this geometry a useful benchmark for comparison with the existing spline-based approach~\citep{shimizu2021construction}. 
The second is a bowl shape (\S\ref{subsec:synthetic2}), an example of non-developable surfaces that intrinsically require stochastic reconstruction for accurate reconstruction. 
This geometry resembles the earthquake fault estimated in our case study of the 2013 Balochistan earthquake (\S\ref{eq:apptoEQ}). 

In both examples, the maximum horizontal component of the normal vector is approximately $n_\perp=0.45$, corresponding to a tilt of approximately 40 degrees from the reference plane. 
These are fairly curved shapes within the range  within tilt angles of 45 degrees, where the conversion from potency beachballs to normal vectors is straightforward. 
Although the maximum tilt is similar in both models, their reconstruction accuracies differ (\S\ref{subsec:synthetic3}).

In the synthetic tests, we assume $\sigma_n<0.25$. 
Because fairly curved surfaces are computed, excessive noise could result in $n^*_z<0$ ($\theta>\pi/2$) in the above definition, which falls outside the scope of this analysis, and even outside the definition range of the hemispherical coordinates. 
In this study, $n^*_z<0$ has been precluded as a subset of $\theta\geq\pi/4$ to guarantee the one-to-one mapping from the potency to the $n$-vector (\S\ref{subsec:forward}), although the $n^*_z<0$ condition is formally accounted for in the loss function of eq.~(\ref{eq:areaelementconversion}) for the use of the spherical coordinates in parametrizing $n$.
Stochastic surface reconstruction involving stochastic $n$-vector selection from nodal planes is left open for future study. 

\subsection{A uniaxial twist}
\label{subsec:synthetic1}

\begin{figure*}
 \includegraphics[width=120mm]{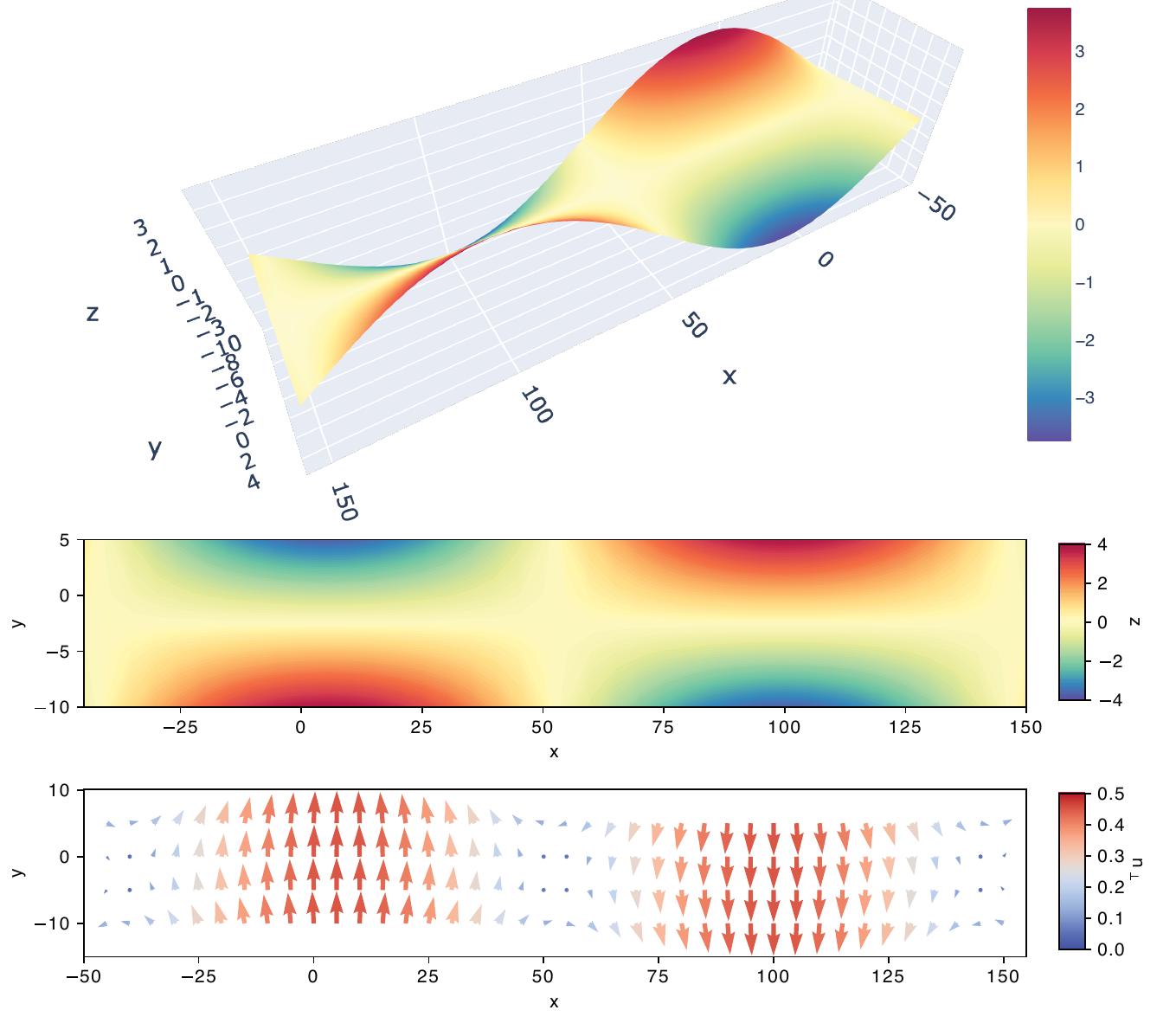}
 \caption{The ground truth of our synthetic test 1: $z=[(y-\bar y)/W]\sin[(x-\bar x)/L]$ with $W=2$ and $L=30$, where $(\bar x,\bar y)$ denotes the reference plane center. 
 }
 \label{fig:synth1}
\end{figure*}

The first case examines the uniaxial twist illustrated in Fig.~\ref{fig:synth1}. Its elevation profile is symmetric with respect to the short axis $y$ (see Fig.~\ref{fig:synth1} caption). Excluding the centerline $y = 0$, where $n_x = 0$ identically, both $n_x$ and $n_y$ are non-zero, resulting in the three-dimensionally varying $n$-vector. Hereafter, the shape is visualized using bird's-eye views.

\begin{figure*}
 \includegraphics[width=120mm]{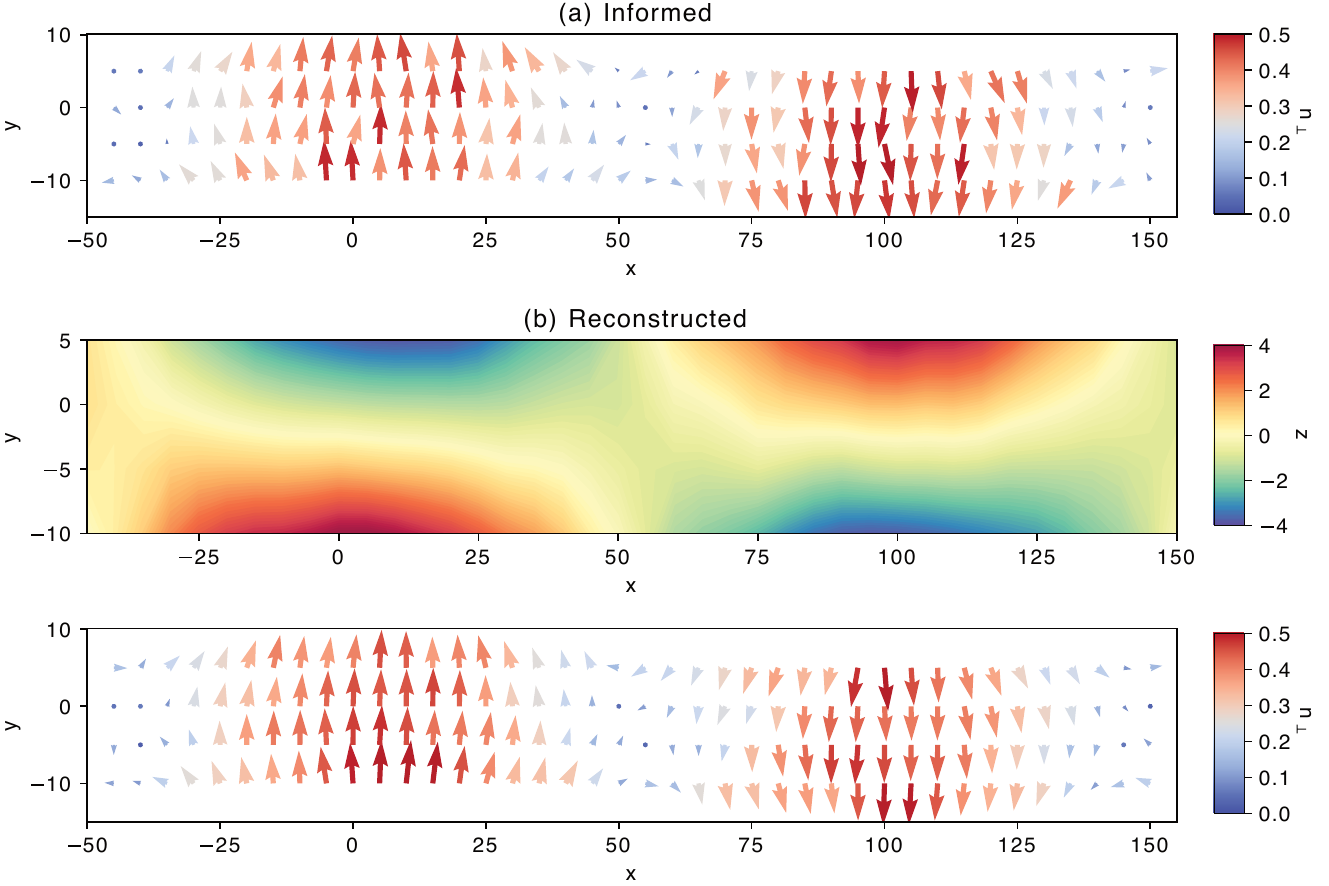}
 \caption{Examples of (a) informed noisy $n$-vectors and (b) reconstructed elevation and $n$-vector fields in our synthetic test 1. The $n$-vector randomly varies from the ground truth with a standard deviation of 0.05. 
 }
 \label{fig:synth2}
\end{figure*}

Fig.~\ref{fig:synth2} shows the estimation results with a noise standard deviation of 0.05. Despite the noise contained in the input $n$-vectors, the estimated elevation is consistent with the original shape. 
Primary geometric features, such as the concavities and convexities at the four corners, are well reproduced, although minor shifts in the contours are observed. 
As noted in the previous section, the random noise component ($\Delta n=n_*-n_0$) almost surely violates the slope integrability condition (eq.~\ref{eq:commutability_constraint_nvect}). Consequently, the most probable surface $\Gamma_*$ (eq.~\ref{eq:optimaloptimal_explicit}) acts as a filter, selectively suppressing the non-integrable noise components in $n_*$.

\subsection{A bowl shape}
\label{subsec:synthetic2}
\begin{figure*}
 \includegraphics[width=120mm]{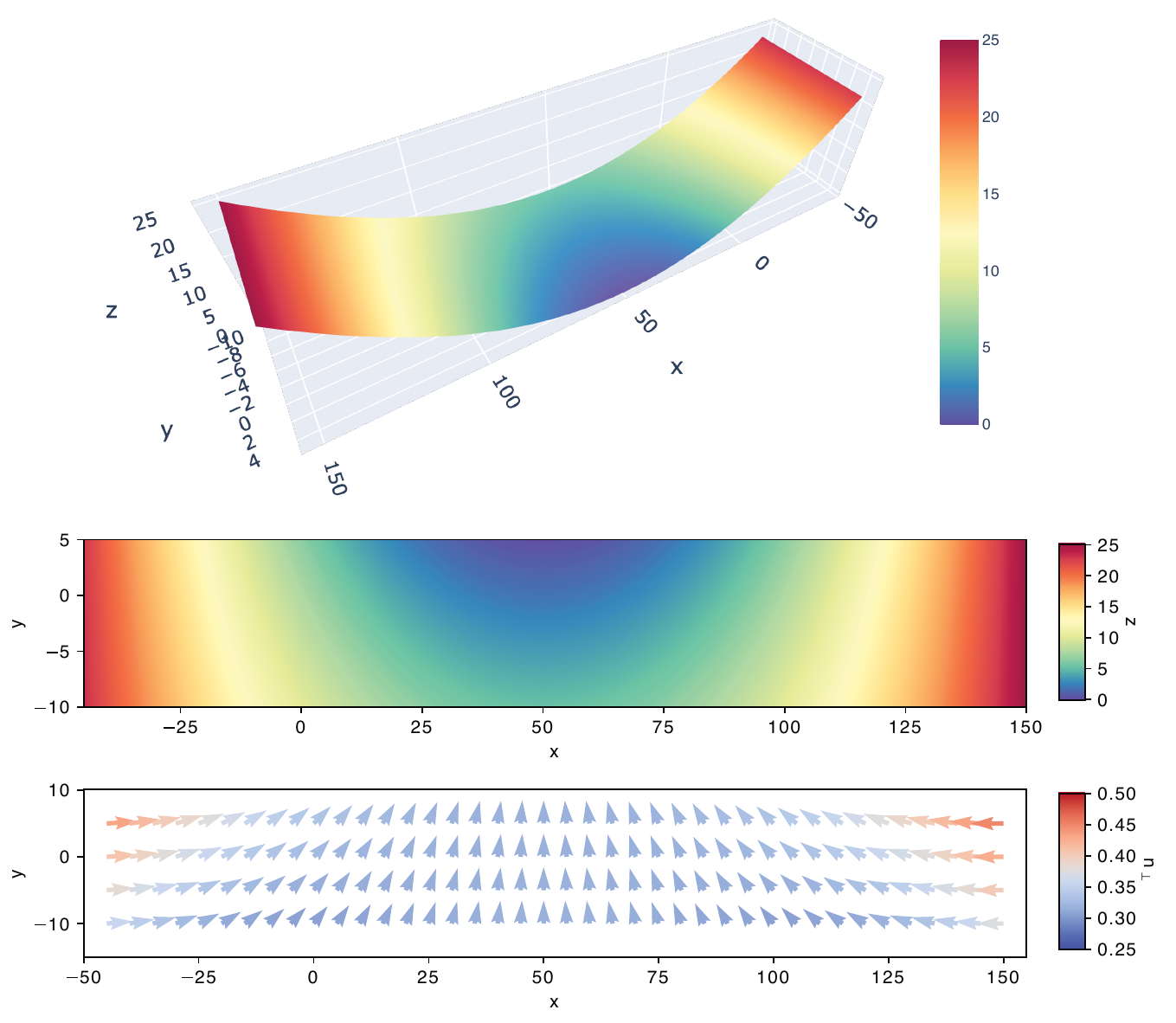}
 \caption{The ground truth of our synthetic test 2: $z=-y/Y_1+(x/X)^2(Y-Y_2)/Y_1$ with $X=50$, $Y_1=3$, and $Y_2/Y_1=-25$. 
 }
 \label{fig:synth3}
\end{figure*}

The second test case examines a bowl-shaped surface illustrated in Fig.~\ref{fig:synth3}, which resembles the geometry estimated for the 2013 Balochistan earthquake (\S\ref{eq:apptoEQ}). 
The elevation $z$ is defined by a function parabolic along the long axis $x$ and linear along the short axis $y$ (see the Fig.~\ref{fig:synth3} caption for details). Interpreting $x$ as the strike axis and $y$ as the dip, the surface exhibits significant curvature near the top and increases vertical tilts toward the lateral end. Qualitatively, larger $|z_{,x}|$ appears for larger $y$ and larger $z_{,y}$ appears for larger $|x|$.

\begin{figure*}
 \includegraphics[width=120mm]{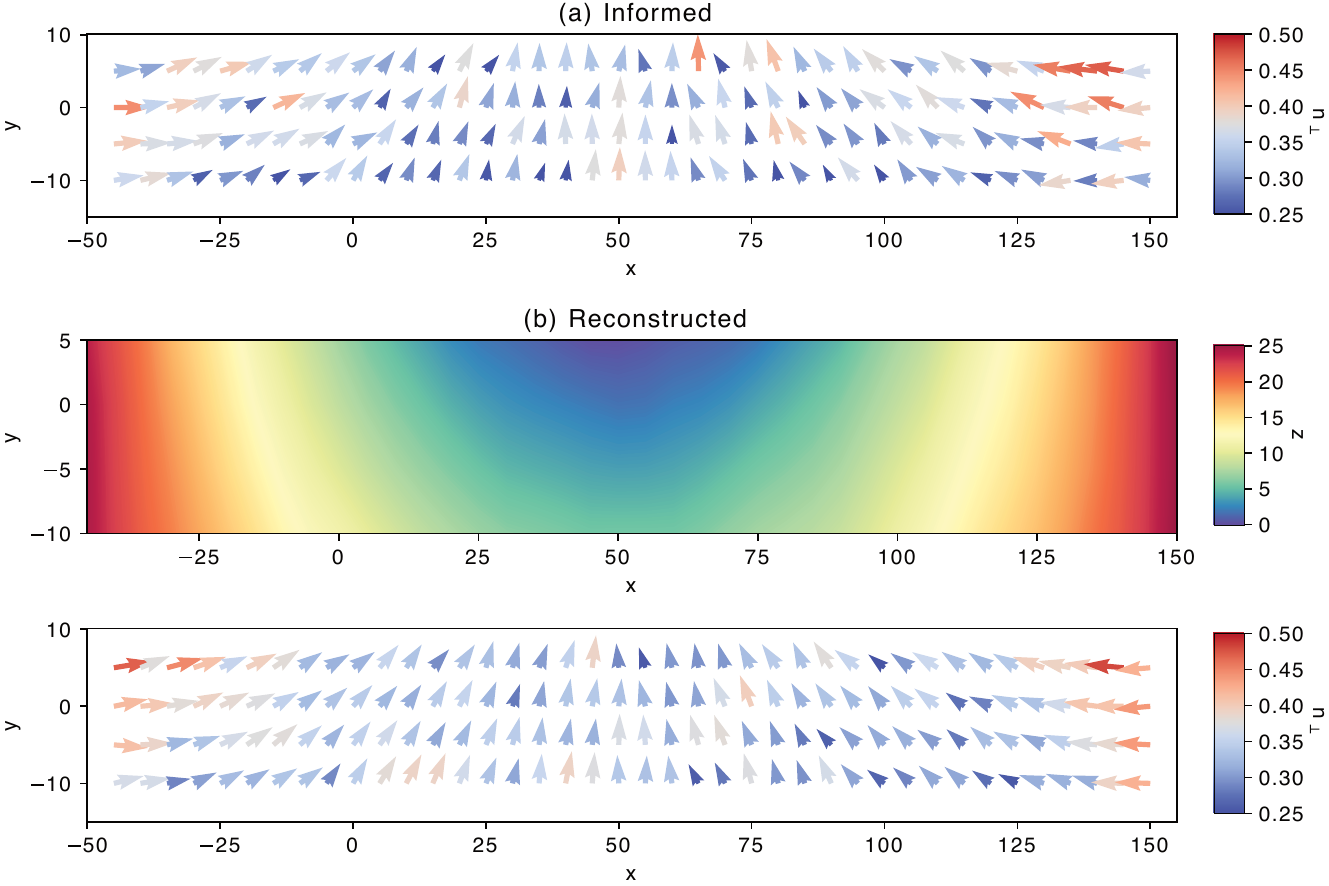}
 \caption{Examples of (a) informed noisy $n$-vectors and (b) reconstructed elevation and $n$-vector fields in our synthetic test 2. The $n$-vector randomly varies from the ground truth with a standard deviation of 0.05. 
 }
 \label{fig:synth4}
\end{figure*}

Fig.~\ref{fig:synth4} shows the results for a standard deviation of 0.05. 
The reconstructed geometry is more stable than the reconstructed uniaxial twist. 
This discrepancy suggests that, for a given level of the $n$-vector noise, short-wavelength features such as topography are more susceptible to noise than global structures. 
Since broad-scale elevations are controlled by long-wavelength slopes, 
we attribute this stability difference to the spatial averaging of integration, which low-passes the slope error in a long-wavelength scale. 
The enhanced surface reproducibility at longer wavelengths might be reinforced in the application to the PDTI, to which the spatial potency smoothing is commonly applied.

\subsection{Robustness tests of surface reconstruction}
\label{subsec:synthetic3}

\begin{figure*}
 \includegraphics[width=120mm]{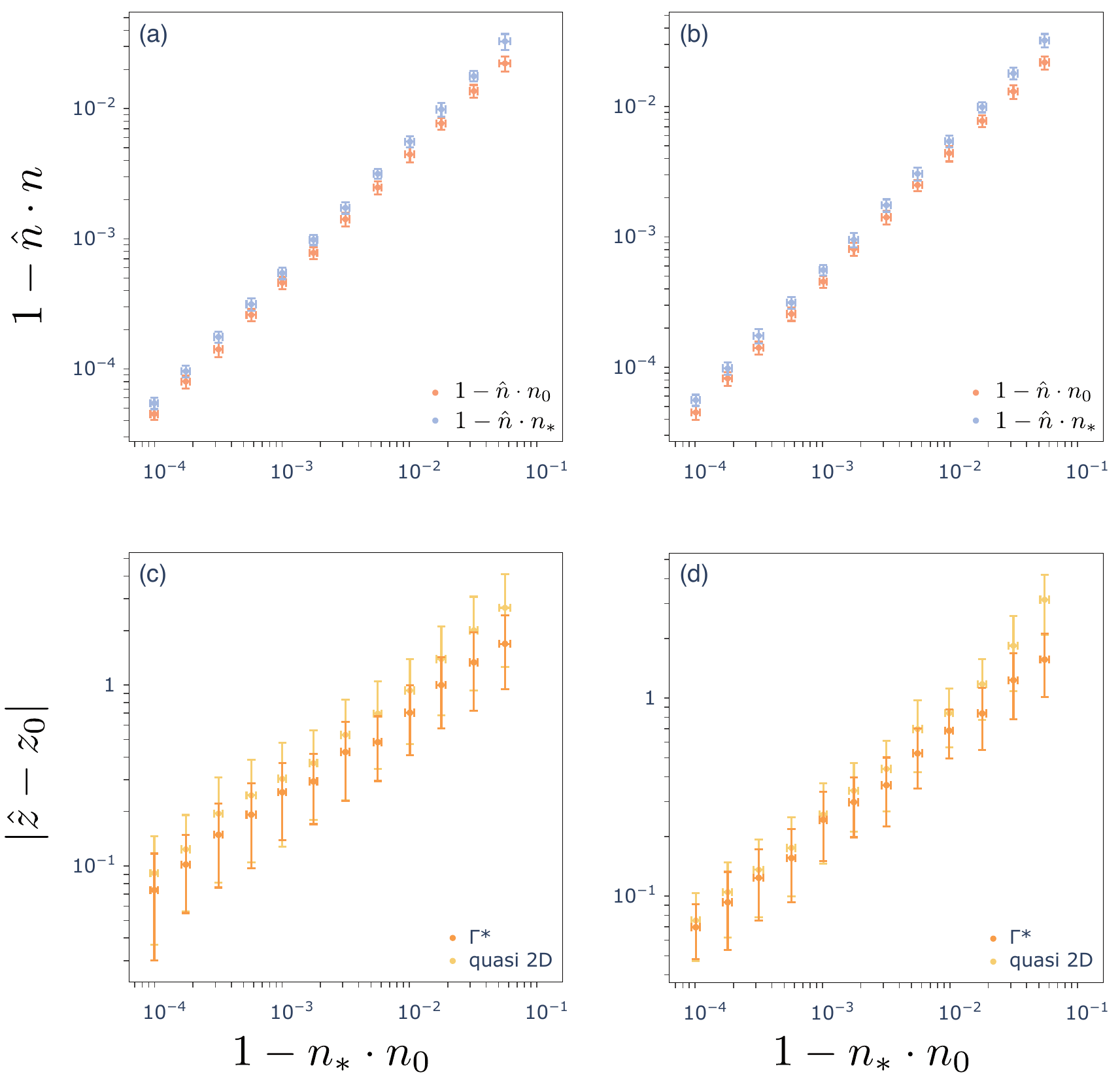}
 \caption{
 Reconstruction errors and residuals in our synthetic tests, plotted in terms of $n$-vectors in tests 1 (a) and 2 (b) and elevations in tests 1 (c) and 2 (d). 
 All plotted values represent spatial averages over the reference plane. The means and standard deviations are derived from 64 independent noise realizations. 
 The horizontal axis $1-n_*\cdot n_0$ represents the noise level of input $n$-vectors $n_*$ relative to the ground truth $n_0$. 
 In the panels (a) and (b), the vertical axis $1-\hat n\cdot n$ represents the errors $1-\hat n\cdot n_0$ (red) and the residuals $1-\hat n\cdot n_*$ (blue) of $n$-vector estimates $\hat n$. 
 In the panels (c) and (d), the vertical axis $|\hat z-z_0|$ represents the error of estimated elevation $\hat z$ relative to the ground truth $z_0$ (orange). 
 Elevation errors of quasi-two-dimensional surface reconstruction (yellow) following \citet{shimizu2021construction} are also indicated for comparison. 
 }
 \label{fig:synth5}
\end{figure*}

Figure~\ref{fig:synth5} presents a summary of the errors and residuals for the two tests described above across varying noise levels.
The estimation errors $1-\hat n\cdot n_0$ and data residuals $1-\hat n\cdot n_*$ for normal vectors (Fig.~\ref{fig:synth5}a, b) both scale linearly with the noise amplitude $1- n_0\cdot n^*$. 
Notably, the data residuals are visibly smaller than the noise amplitude, indicating significant variance reduction. 
The estimation error $1-\hat n\cdot n_0$ is smaller than the data residual $1-\hat n\cdot n_*$. This suggests that the reconstruction process does not merely explain the noisy input $n$-vectors but rather effectively denoises them. 

The spatial average of elevation errors $|z-\bar z|$ ($\Gamma_*$ in Fig.~\ref{fig:synth5}c, d) exhibits larger scatter compared to that of normal vectors (Fig.~\ref{fig:synth5}a, b). Its average is roughly proportional to the square root of the noise amplitude of the normal vectors $1-n_0\cdot n^*$, which approximate squared angle fluctuations, implying a linear proportionality between the input and output standard deviations. 
The elevation residuals are roughly the same between the two shapes (Fig.~\ref{fig:synth5}c, d). 
Those same-order absolute residuals indicate a significant difference in relative residuals, supporting superior reproducibility of a global curve (bowl-shape) compared to local variations (uniaxial twist) observed in the previous subsections.

For comparison, we also evaluated quasi-two-dimensional surface construction from \citet{shimizu2021construction} (quasi 2D in Fig.~\ref{fig:synth5}c, d). 
This method approximates the elevation as $z(x,y)\simeq z(\bar x,\bar y)+\int_{\bar x}^x dx^\prime \bar z_{,x}(x^\prime)+\bar z_{,y}(x)(y-\bar y)$ by using the mean slope $\bar z_{,\alpha}=-\int d_y (n^*_\alpha/n^*_z)/\int dy$ integrated along the short axis $y$. In this section, we assume $z(\bar x,\bar y)$ at the reference plane center $(\bar x,\bar y)$ is known to compute the quasi-two-dimensional reconstruction. 
The proposed stochastic surface reconstruction $\Gamma_*$ performs comparably to quasi-two-dimensional reconstruction at low noise levels. However, $\Gamma_*$ yields smaller estimation errors at high-noise levels. Even for a uniaxial twist, for which quasi-two-dimensional approximation is errorless, the accuracy of  $\Gamma_*$ is higher, although the difference is within accuracy variations (Fig.~\ref{fig:synth5}c). 
For a bowl shape which resembles the fault shape inferred in our application (\S\ref{eq:apptoEQ}), the accuracy of $\Gamma_*$ is several times higher than the quasi-two-dimensional benchmark for the maximum noise amplitude tested (Fig.~\ref{fig:synth5}d). 

The fault models used here have high aspect ratios, a condition where quasi-two-dimensional interpolation from \citet{shimizu2021construction} typically provides a good approximation in the absence of noise. 
Even for such narrow faults, the above experiments indicate that 
the probabilistic approach offers significant advantages for surface reconstruction from noisy $n$-vector fields. 
Rather than reducing noise through the law of large numbers associated with averaging along the short axis, denoising based on the a priori knowledge that $n$-vector components that violate the slope integrability condition (eq.~\ref{eq:commutability_constraint_nvect}) are fictitious is likely to better reproduce the original shape.

\section{Application}
\label{sec:app}
Finally, to validate the practical applicability of our approach, we incorporate the proposed normal-to-elevation projection into the surface reconstruction from the PDTI~\citep{shimizu2021construction} and apply it to the 2013 Balochistan earthquake. This case study serves to demonstrate the method's capability to reconstruct three-dimensional fault geometry partly unconstrained in previous efforts. 

\subsection{Incorporating three-dimensional surface reconstruction with Bayesian PDTI}
\label{sec:probabilisticShimizu2021}

For this application, we integrate the prior constraint on fault geometry $P(\Gamma|\mathbf{a})$ with the posterior probability of the potency density derived from the orthodox PDTI framework. 
While we have focused on static problems thus far, we now extend the formulation to dynamic problems for inverting teleseismic data. 
The formulation derived in previous sections remains valid even when the representation theorem becomes time-dependent and the basis functions span spatiotemporal domains. Time-dependent geometry can be also incorporated, but in this study, we focus on the time-invariant geometry. 

The formulation of the orthodox PDTI, for both static and dynamic problems, comprises 
the potency likelihood $P({\bf d}|{\bf a},\Gamma;\boldsymbol\sigma^2)$, which is given by data ${\bf d}$, and the prior $P({\bf a};\boldsymbol\rho^2)$ for the potency density parameterized by basis function expansions ${\bf a}$. 
The scales of variances for the likelihood $\boldsymbol\sigma^2$ 
and weight factors for prior loss functions $\boldsymbol\rho^2$ are simultaneously estimated. Usually, $\boldsymbol\sigma^2$ represents the scales of observation errors and Green's function errors~\citep{yagi2011introduction}, and $\boldsymbol\rho^2$ represents the smoothing intensities with respect to space and time~\citep{shimizu2020development}. 
The problem to be solved here consists of 
these $P({\bf d}|{\bf a},\Gamma;\boldsymbol\sigma^2)$ and $P({\bf a};\boldsymbol\rho^2)$ 
and 
the prior for the surface $P(\Gamma|{\bf a})$. Its formal solution is the joint posterior of ${\bf a}$ and $\Gamma$:
\begin{equation}
P({\bf a},\Gamma|{\bf d};\boldsymbol\sigma^2,\boldsymbol\rho^2)=P({\bf d}|{\bf a},\Gamma;\boldsymbol\sigma^2)
P({\bf a};\boldsymbol\rho^2)P(\Gamma|{\bf a})/P({\bf d};\boldsymbol\sigma^2,\boldsymbol\rho^2),
\end{equation}
where
\begin{equation}
    P({\bf d};\boldsymbol\sigma^2,\boldsymbol\rho^2)=\int d{\bf a}d\Gamma P({\bf d}|{\bf a},\Gamma;\boldsymbol\sigma^2)
P({\bf a};\boldsymbol\rho^2)P(\Gamma|{\bf a}).
\label{eq:ABIC}
\end{equation}

The analysis for the dislocation limit is simplified when the joint posterior $P({\bf a},\Gamma|{\bf d};\boldsymbol\sigma^2,\boldsymbol\rho^2)$ 
is decomposed into the 
marginal posterior of the potency $P({\bf a}|{\bf d};{\boldsymbol\sigma}^2,{\boldsymbol\rho}^2)=\int d\Gamma P({\bf a},\Gamma|{\bf d}; {\boldsymbol\sigma}^2,{\boldsymbol\rho}^2)$
and the conditional posterior of the surface $P(\Gamma|{\bf a}, {\bf d};{\boldsymbol\sigma}^2,{\boldsymbol\rho}^2)$: 
\begin{equation}
   P({\bf a},\Gamma|{\bf d};{\boldsymbol\sigma}^2,{\boldsymbol\rho}^2)=P({\bf a}|{\bf d};{\boldsymbol\sigma}^2,{\boldsymbol\rho}^2)P(\Gamma|{\bf a}, {\bf d};{\boldsymbol\sigma}^2,{\boldsymbol\rho}^2).
   \label{eq:aGamma_separation}
\end{equation}
From eq.~(\ref{eq:prior_Gamma_limit}), \begin{flalign}
    P({\bf a}|{\bf d};{\boldsymbol\sigma}^2,{\boldsymbol\rho}^2)&=P({\bf a},\Gamma_*({\bf a})|{\bf d};{\boldsymbol\sigma}^2,{\boldsymbol\rho}^2)
    \label{eq:a_marginal}
    \\
    P(\Gamma|{\bf a}, {\bf d};{\boldsymbol\sigma}^2,{\boldsymbol\rho}^2)&=\delta(\Gamma-\Gamma_*({\bf a})).
    \label{eq:Gamma_conditional}
\end{flalign}
Namely, the evaluation of $P({\bf a},\Gamma|{\bf d};\boldsymbol\sigma^2,\boldsymbol\rho^2)$ reduces to an alternating recursion of 
the most probable estimation of $\Gamma$ given ${\bf a}$, which returns $\Gamma_*$ (eq.~\ref{eq:Gamma_conditional}),
and 
the inference of ${\bf a}$ given $\Gamma_*$ (eq.~\ref{eq:a_marginal}).  

We evaluate the hyperparameter optimals $\hat {\boldsymbol\sigma}^2$ and $\hat{\boldsymbol\rho}^2$ by Akaike's Bayesian Information Criterion~\citep[ABIC;][]{akaike1980likelihood}: 
\begin{equation}
    (\hat {\boldsymbol\sigma}^2,\hat{\boldsymbol\rho}^2)={\rm argmax}_{{\boldsymbol\sigma}^2,{\boldsymbol\rho}^2} P({\bf d};{\boldsymbol\sigma}^2,{\boldsymbol\rho}^2).
    \label{eq:optimumh}
\end{equation}
The associated optimals of the potency density and fault surface are evaluated by the posterior maximum: 
\begin{equation}
    (\hat {\bf a},\hat \Gamma)={\rm argmax}_{{\bf a},\Gamma}P({\bf a},\Gamma| {\bf d};\hat {\boldsymbol\sigma}^2,\hat{\boldsymbol\rho}^2).
    \label{eq:optimumaGamma}
\end{equation}

Neglecting third- and higher-order moments from the probability peaks as in \citet{shimizu2021construction}, 
eqs.~(\ref{eq:ABIC}) and (\ref{eq:a_marginal}) are simplified as
\begin{equation}
    P({\bf d};{\boldsymbol\sigma}^2,{\boldsymbol\rho}^2)\simeq \int d{\bf a} P({\bf d}|{\bf a},\hat \Gamma; {\boldsymbol\sigma}^2)
P({\bf a};{\boldsymbol\rho}^2)
\label{eq:ABICapprox}
\end{equation}
and
\begin{equation}
    P({\bf a}|{\bf d};{\boldsymbol\sigma}^2,{\boldsymbol\rho}^2)\simeq \frac{P({\bf d}|{\bf a},\hat \Gamma;{\boldsymbol\sigma}^2)
P({\bf a};{\boldsymbol\rho}^2)}{P({\bf d};{\boldsymbol\sigma}^2,{\boldsymbol\rho}^2)}.
\label{eq:reducedmarginalposteriorofa}
\end{equation}
Besides, from eqs.~(\ref{eq:aGamma_separation}) and (\ref{eq:Gamma_conditional}), 
\begin{equation}
    \hat \Gamma=\Gamma_*(\hat {\bf a}).
    \label{eq:Gammahat}
\end{equation}
Thus, our optimum-value estimation is an alternating recursion: solving for $\hat\Gamma$ given $\hat {\bf a}$ and 
solving for $(\hat {\bf a},\hat {\boldsymbol\sigma}^2,\hat {\boldsymbol\rho}^2)$ given $\hat \Gamma$. 
The method of analysis with a fixed $\Gamma$ corresponds exactly to the orthodox source inversion with fixed geometry~\citep{yagi2011introduction,shimizu2020development}, 
and the analysis of $\Gamma$ given ${\bf a}$ represents the geometry reconstruction we have formulated. 
Once 
$\hat \Gamma$ and $(\hat {\boldsymbol\sigma}^2,\hat {\boldsymbol\rho}^2)$ are obtained, 
the uncertainty of ${\bf a}$ can be evaluated through eq.~(\ref{eq:reducedmarginalposteriorofa}) using $P({\bf a}|{\bf d};\hat {\boldsymbol\sigma}^2,\hat {\boldsymbol\rho}^2)$. 
Hyperparameter uncertainty is similarly evaluable using $P({\bf d};{\boldsymbol\sigma}^2,{\boldsymbol\rho}^2)$ for given $\hat\Gamma$, although this distribution is a sharply peaked and well-behaved, following the law of large numbers exceptionally rapidly when ${\boldsymbol\sigma}^2$ and ${\boldsymbol\rho}^2$ are exponential loss weights~\citep{sato2022appropriate}. 

The above procedure using $P({\bf a}|{\bf d};{\boldsymbol\sigma}^2,{\boldsymbol\rho}^2)$ and $P({\bf d};{\boldsymbol\sigma}^2,{\boldsymbol\rho}^2)$ does not explicitly evaluate $\Gamma$ uncertainty. 
In the present problem setting using a delta-functional $P(\Gamma|{\bf a})$, 
the estimation error of $\Gamma$ solely derives from the error propagation of the potency density estimation. 
Its evaluation requires another joint posterior decomposition $
   P({\bf a},\Gamma|{\bf d};{\boldsymbol\sigma}^2,{\boldsymbol\rho}^2)=P(\Gamma|{\bf d};{\boldsymbol\sigma}^2,{\boldsymbol\rho}^2)P({\bf a}|\Gamma, {\bf d};{\boldsymbol\sigma}^2,{\boldsymbol\rho}^2)
$ and the evaluation of $P(\Gamma|{\bf d};{\boldsymbol\sigma}^2,{\boldsymbol\rho}^2)=\int d{\bf a}P({\bf a},\Gamma|{\bf d};{\boldsymbol\sigma}^2,{\boldsymbol\rho}^2)$. 
Due to the slope integrability (eq.~\ref{eq:commutability_constraint_nvect}) on a smooth surface, 
the posterior of the fault surface $P(\Gamma|{\bf d};{\boldsymbol\sigma}^2,{\boldsymbol\rho}^2)$ is distinct from the posterior of $n_*$, the $n$-vector field expected from the potency density (Appendix \ref{app:A}).
The evaluation of $P({\bf a}|\Gamma, {\bf d};{\boldsymbol\sigma}^2,{\boldsymbol\rho}^2)$ corresponds to the conventional slip inversion, where the potency density ${\bf a}$ is constrained to represent a displacement discontinuity field across given $\Gamma$. 

\subsection{Recipe}
\label{sec:recipe}

The computations of $\hat {\bf a}$, $\hat \Gamma$, and $(\hat{\boldsymbol\sigma}^2,\hat{\boldsymbol\rho}^2)$ (eqs.~\ref{eq:optimumh} and \ref{eq:optimumaGamma}), as implemented in eqs.~(\ref{eq:ABICapprox})--(\ref{eq:Gammahat}), are formally identical to
those of \citet{shimizu2021construction}, which alternately evaluate $({\bf a},{\boldsymbol\sigma}^2,{\boldsymbol\rho}^2)$ for a given $\Gamma$ and $\Gamma$ for a given ${\bf a}$. Consequently, 
simply updating the fault reconstruction stage of their routine
enables the reconstruction of three-dimensional fault geometry. 

\begin{figure*}
 \includegraphics[width=130mm]{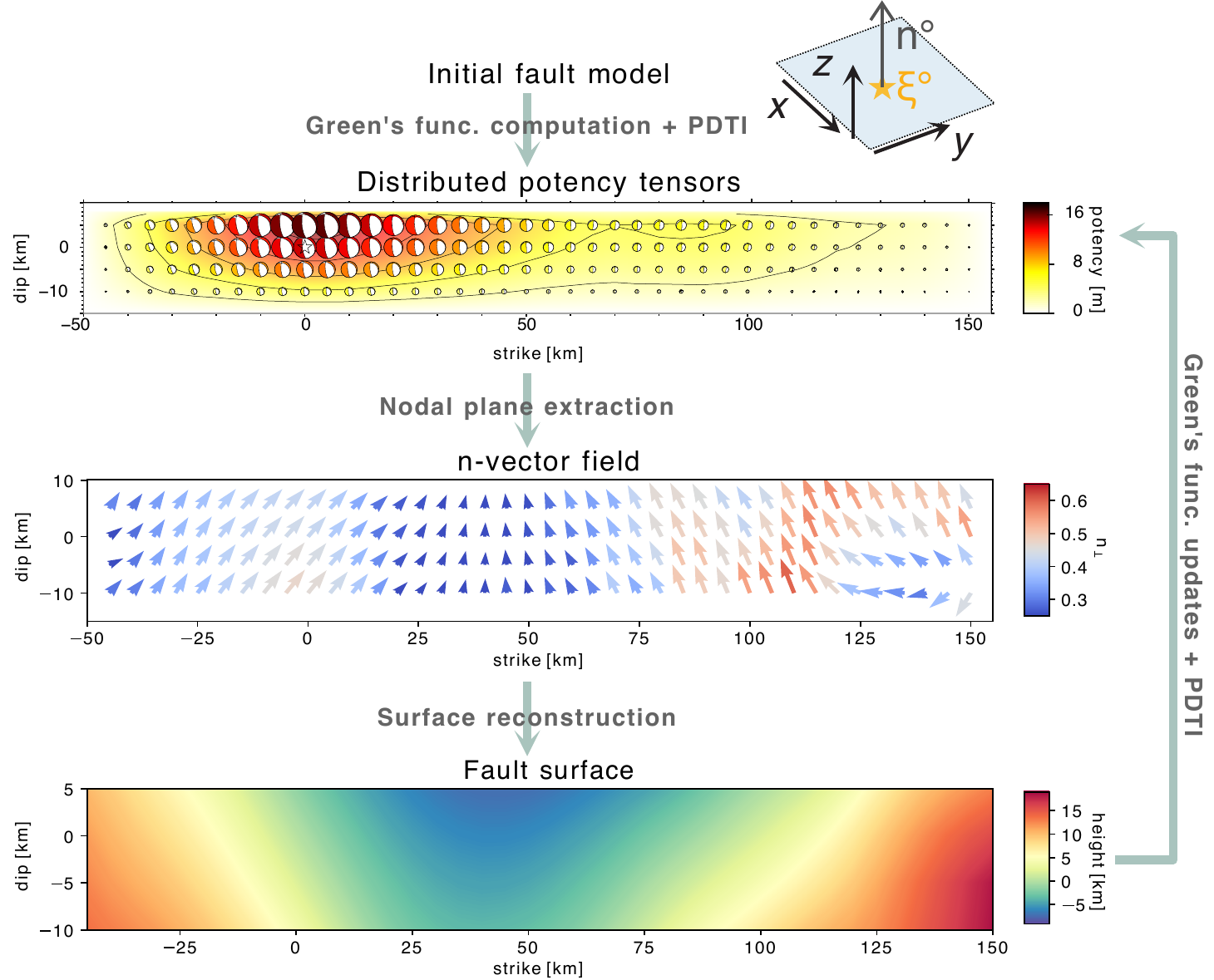}
 \caption{
 A schematic diagram of simultaneous potency and fault shape estimation as per eq.~(\ref{eq:iteration_3dfaultinfPDTI}). The loop starts with the fault model initialized. 
 Distributed potency tensors obtained via the PDTI are converted into an $n$-vector field, which reconstructs the fault surface. 
 The reconstructed surface updates Green's functions, and alternate potency density and shape inferences restart. 
 The distributed potency-tensor solutions in the figure represent the estimate of the first iteration obtained by \citet{shimizu2020development}, and the other elements refer to the results of Fig.~\ref{fig:app1}. 
 }
 \label{fig:5}
\end{figure*}

Figure~\ref{fig:5} outlines our computational routine that follows \citet{shimizu2021construction}. 
(I: Fault model initialization) A structured grid is defined along the reference plane, with the spatial knots for potency density interpolations placed at grid centers; nonplanar initializations following stage V can be optionally applied. 
(II: PDTI with Green's function computations) The PDTI is performed with Green's function approximated by point-source representations; when the point sources are allocated to grid centers, they are weighted by the areas of gridded subsurfaces. 
(III: Nodal plane extraction) 
The $n$-vector estimates for point sources are computed as the nodal planes of estimated potency beachballs; the $n$-vector field estimate is approximated by their interpolation for tractability. 
(IV: Surface reconstruction) The elevation field is reconstructed from the $n$-vector field estimate. 
(V: PDTI with Green's function updates) The structured grid is vertically projected onto the reconstructed fault surface, and the vertical positions of sources and the areas of gridded subsurfaces are updated accordingly; the PDTI then updates the potency beachballs, and the loop returns to the stage III.

In summary, once the PDTI is performed, the potency-tensor solutions set an $n$-vector field, which in turn reconstructs the fault surface. This continuous surface is subdivided as a digital elevation model (DEM) to update Green's function for the subsequent PDTI iteration. 
We implement the DEM operation by the Visualization Toolkit~\citep{schroeder1998visualization} via a Python interface PyVista~\citep[][]{sullivan2019pyvista}. 
PyVista is utilized for the following two operations: (A) 
projecting the structured grid, which is generated by numpy.meshgrid, onto the reconstructed fault surface (pyvista.StructuredGrid)
and 
(B) computing source locations and subdivided areas on the vertically projected structured grids (compute\_cell\_sizes, cell\_centers). 

We solve this recursion by a gradient descent, where the $\Gamma_*$ determination (eq.~\ref{eq:optimaloptimal_explicit}) is modified to include a penalty term on the residual elevation relative to a given shape $\bar\Gamma$ with elevation $\bar z$, \begin{equation}
\begin{aligned}    
    \Gamma_*^\prime({\bf a},\bar \Gamma;\lambda)={\rm argmin}\left\{\int
n_{*z}( n_{*\parallel x}n_{*z} z_{,x}+n_{*\parallel y}n_{*z} z_{,y}+n_{*\perp})^2 dxdy
\right.
\\
\left.
+
\int 
n_{*z}(n_{*\parallel x} z_{,y}- n_{*\parallel y} z_{,x})
^2dxdy+\lambda^{-1}\int |z-\bar z|^2 dxdy
\right\},
\end{aligned}
    \label{eq:defofGammastar_prime}
\end{equation}
and each surface update accounts for the pre-update surface:
\begin{equation}
    \begin{aligned}        
    (\hat{\boldsymbol\sigma}^{2(n)},\hat{\boldsymbol\rho}^{2(n)})&:={\rm argmax}_{({\boldsymbol\sigma}^{2},{\boldsymbol\rho}^{2})}
\int d{\bf a}  P({\bf d}|{\bf a},\hat \Gamma^{(n)}; \boldsymbol\sigma^2)
P({\bf a};{\boldsymbol\rho}^2)
    \\
    \hat{\bf a}^{(n)}&:={\rm argmax}_{\bf a}
    P({\bf d}|{\bf a},\hat \Gamma^{(n)};\hat {\boldsymbol\sigma}^{2(n)})
P({\bf a};\hat {\boldsymbol\rho}^{2(n)})
    \\
    \hat \Gamma^{(n+1)}&:=\Gamma_*^\prime(\hat {\bf a}^{(n)},\hat \Gamma^{(n)};\lambda).
    \end{aligned}
    \label{eq:iteration_3dfaultinfPDTI}
\end{equation}
The elevation $\hat z^{(n)}$ of the $n$-th iteration $\hat \Gamma^{(n)}$ 
is updated to $\hat z^{(n+1)}=\hat z^{(n)}+\lambda \partial [\kappa^{-1}\ln P(\Gamma| \hat {\bf a}^{(n)})]/\partial z$, where $\lambda$ serves as the learning rate of this gradient descent. 
When iterations converge, $\hat z^{(n)}=\hat z^{(n+1)}$ is met, 
and the surface optimal $\hat \Gamma=\hat \Gamma^{(n+1)}$ is the solution that satisfies $\partial\ln P(\Gamma| \hat {\bf a})/\partial z= 0$
for $\hat {\bf a}=\hat {\bf a}^{(n)}$.

Because the prior information provided by eq.~(\ref{eq:prior_Gamma}) is set independently of reference-surface coordinates, the formulation and the resultant posterior is reference-invariant. However, this reference-surface independence does not strictly hold for the numerical implementation. 
Numerical implementation necessitates a finite computational domain, which implicitly constrains the spatial extent of the source region and potentially influences the estimation, a situation analogous to slip inversion. Moreover, since the posterior distribution may exhibit multiple local minima, the surface set as the initial value for the iteration can affect the minimum to which the estimation converges; the initial fault model is set as a reference plane here, although it need not be the reference plane. Note that the hyperparameter inference involved here is nonlinear even in linear problems; proofs of unimodality in hyperparameter estimation exist only under specific conditions~\citep[e.g.,][]{sato2022proof}, to our knowledge. The combination of two linear inverse problems constitutes an overall non-linear inverse problem, for which the posterior unimodality is not guaranteed. 
At present, the fault surface reconstruction from the PDTI described in eq.~(\ref{eq:iteration_3dfaultinfPDTI}) is a purely heuristic approach.
This initial-value dependence characterizes our numerical method as a perturbative analysis refining the initial model, specifically with respect to the source-receiver distance which permits the point-source approximation. Here, the initial fault model, essentially referring to a point-source solution, serves as the base state for perturbation, and the proposed recipe functions as a viable shape refinement of this base geometry.

\subsection{Case study of the 2013 Balochistan earthquake}
\label{eq:apptoEQ}
The 2013 Balochistan earthquake struck southwestern Pakistan on 24 September 2013.  
It was a moment magnitude (Mw) 7.7 event characterized by oblique strike-slip faulting. Variations in the centroid moment tensor solutions and geological features~\citep[e.g.][]{avouac20142013,jolivet20142013} may collectively suggest source complexities typified by fault nonplanarity.
The slip zone corresponds to the Hoshab fault, situated in the transition between the Chaman fault system and the accretionary wedge~\citep{avouac20142013}, suggesting a wedge-slicing listric fault geometry. However, previous multiple point-source approaches struggled to constrain whether a listric or non-listric geometry is preferred for this earthquake~\citep{jolivet20142013}. 
While the PDTI detects major along-strike variations of potency tensors and minor along-dip variations in potency-density tensors, three-dimensional geometry construction remained beyond the scope of previous efforts~\citep{shimizu2020development,shimizu2021construction}. 
By extending the PDTI to reconstruct three-dimensional geometry, we aim to extract the listric geometry of this earthquake fault.

The sole modification from the method and data of \citet{shimizu2021construction} is the implementation of the three-dimensional fault reconstruction described in eq.~(\ref{eq:iteration_3dfaultinfPDTI}). The developed fault reconstruction method is used with the parametrization of the fault elevation adopted in our numerical experiments. 
The fault elevation is expressed by the quadratic B-spline functions as in the PDTI, with the knot interval 1.5 times wider than those of the potency density tensor. The surface integral of the loss function to reconstruct the fault surface is collocated at the $x$-$y$ positions of the spline knots set in the PDTI. The data and processing procedure in this application follow those of \citet{shimizu2020development}, utilizing
observed vertical components of
teleseismic P waveforms converted to velocities at 36 stations shown in Figs.~2(c) and S1 of \citet{shimizu2021construction}.

\begin{figure*}
 \includegraphics[width=135mm]{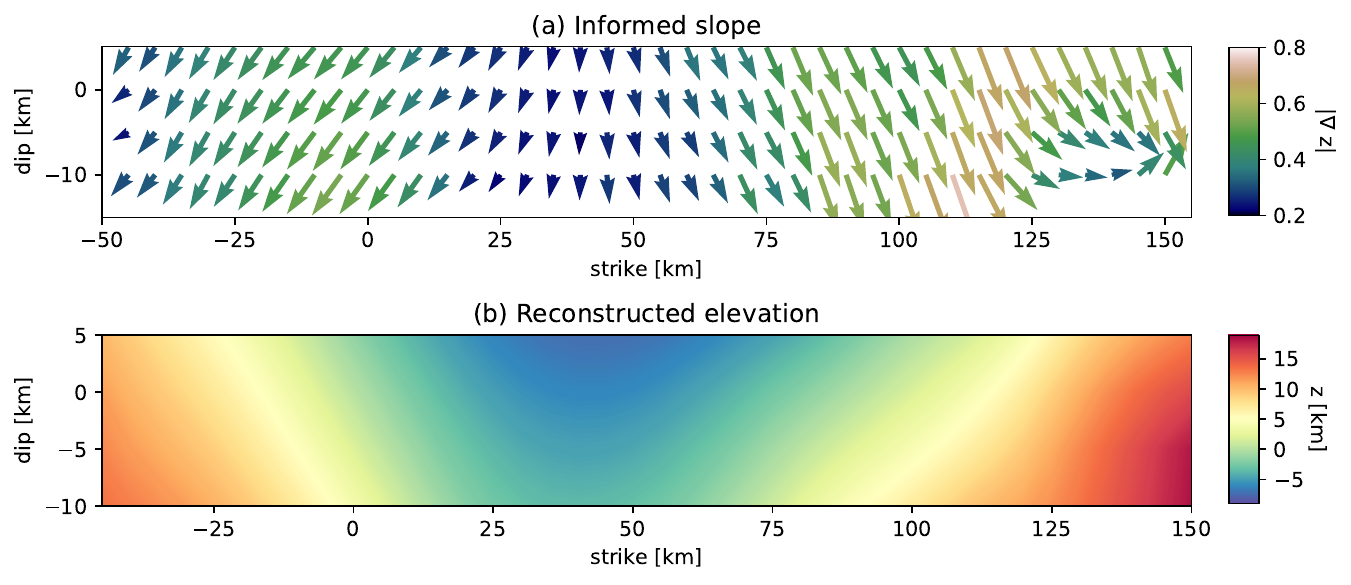}
 \caption{
 Surface reconstruction from the potency tensors distributed over a reference plane. 
 (a) Informed fault slopes, calculated from the estimated $n$-vector field via eq.~(\ref{eq:elevationgradientvsnormal}). Collocation points of informed slopes correspond to the $x$-$y$ locations of B-spline knots for potency densities.
 (b) Reconstructed fault elevation when using eq.~(\ref{eq:optimaloptimal_explicit}). 
}
 \label{fig:app1}
\end{figure*}

\begin{figure*}
 \includegraphics[width=130mm]{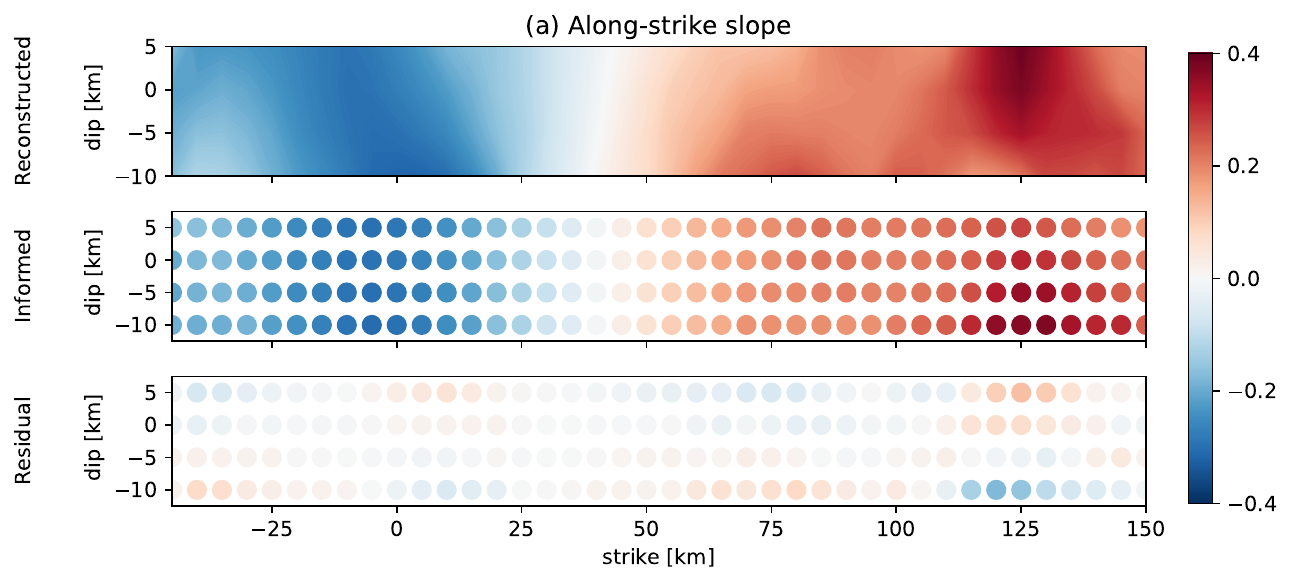}\\
 \includegraphics[width=130mm]{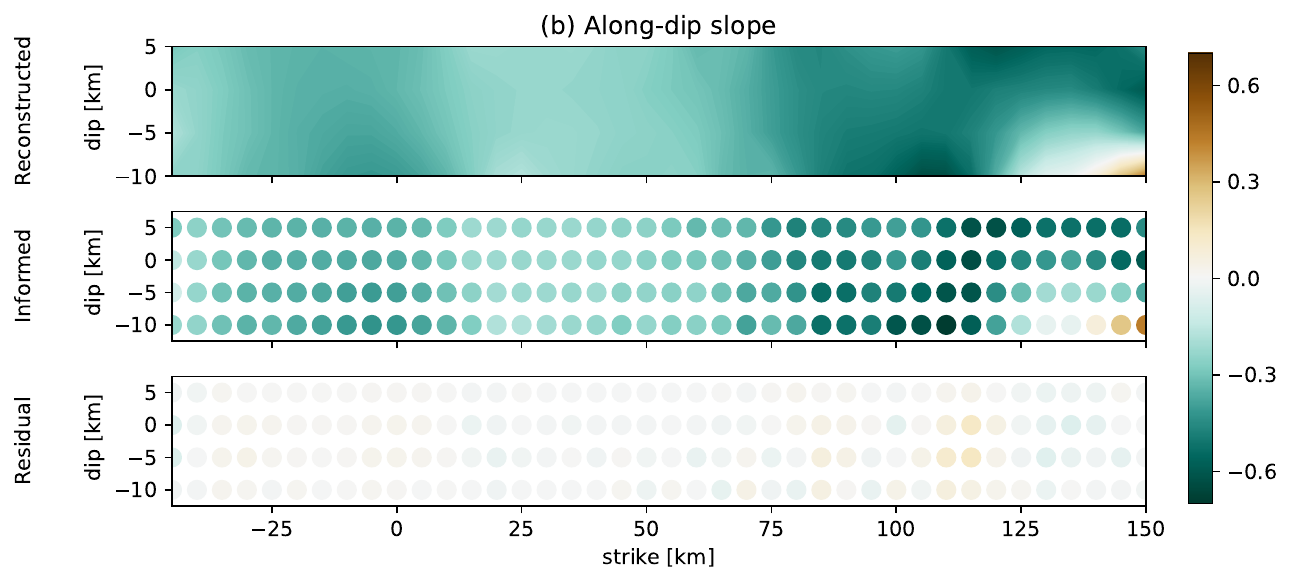}
 \caption{
 Comparisons between reconstructed and informed slopes along strike (a) and dip (b). Reconstructed and informed slopes correspond to Fig.~\ref{fig:app1}b and Fig.~\ref{fig:app1}a, respectively. Their residuals are also indicated. 
}
 \label{fig:app2}
\end{figure*}

Since previous studies have shown that geometry updates largely converge within the first iteration~\citep{shimizu2021construction}, it is instructive to examine the initial surface reconstruction. Figure~\ref{fig:app1} shows the surface reconstruction from the potency-tensor solutions distributed across the reference plane~\citep{shimizu2020development}, which is the initial input of the surface reconstruction also in \citet{shimizu2021construction}. The reference plane is set as a vertical plane passing through the hypocenter with a strike angle of 226 degrees, with the southwest and the upward directions defined as positive. 
The inference shown in Fig.~\ref{fig:app1} employs $\lambda^{-1}=0$ and uses eq.~(\ref{eq:optimaloptimal_explicit}) as done in the synthetic tests. 
Normal vectors evaluated from the PDTI provide expected fault slopes as an input set (Fig.~\ref{fig:app1}a), from which the fault elevation field is reconstructed as the output (Fig.~\ref{fig:app1}b). The slopes of the reconstructed surface explain the input slopes well (Fig.~\ref{fig:app2}). Residuals of $n$-vectors are generally random, although they exhibit some spatial correlation with neighboring points largely due to the smoothing imposed on the potency density estimate. 
Subdividing the smooth surface reconstruction provides subfault elevations and subfault areas, which convert the areal density of potency to distributed potency sources to update Green's function for the PDTI of the next iteration.

\begin{figure*} 
 \includegraphics[width=135mm]{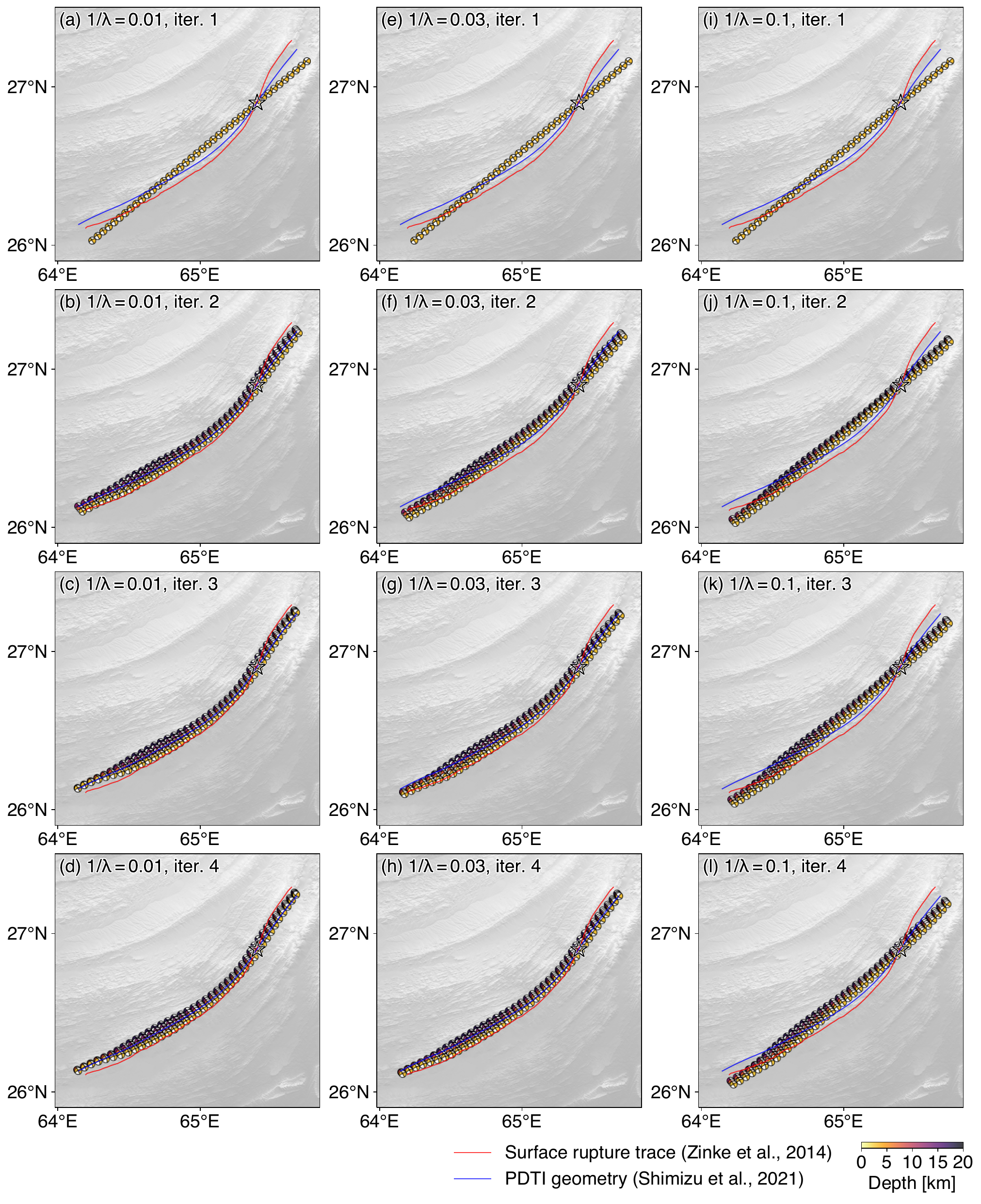}
 \caption{
 Iterations of computed gradient descents. The first to fourth iterations are shown for the learning rate inverse $\lambda^{-1}$ of 0.01 (a--d), 0.03 (e--h), and 0.1 (i--l). 
 The star shows the epicenter, and the background topography is from SRTMGL1~\citep{NASAJPL2013}. 
 The surface rupture trace~\citep[red]{zinke2014surface} and two-dimensional geometry reconstructed in the previous PDTI \citep[blue]{shimizu2021construction} are drawn for comparison.
 }
 \label{fig:app3}
\end{figure*}

Fig.~\ref{fig:app3} shows iterative shape updates. 
The fault geometry is rapidly constrained in areas of large slips, whereas the convergence is slow at the edges characterized by small slips, where fault shape is less constrained. 
High learning rates cause oscillation in poorly constrained regions (fault edges in Fig.~\ref{fig:app3}a--d) while low learning rates slow down updates (Fig.~\ref{fig:app3}i--l), a common characteristic of gradient descents. 
Inference with adjusted learning rates converges to the optimal within a few iterations (Fig.~\ref{fig:app3}e--h). 
The initial surface reconstruction of relatively high learning rates (Fig.~\ref{fig:app3}b) already outlines a converged fault shape, as in a quasi-two-dimensional approach by \citet{shimizu2021construction}. The estimated fault shape progressively approaches the surface rupture identified in the field survey~\citep[Fig.~\ref{fig:app3} red line]{zinke2014surface}, despite the absence of surface trace constraints. In \citet{shimizu2021construction}, the dip was fixed at 90 degrees, causing their estimate (Fig.~\ref{fig:app3}, blue line) to shift northward from the surface trace approximately 50 km southwest of the hypocenter.  This discrepancy is resolved in the present inversion.

\begin{figure*}
 \includegraphics[width=135mm]{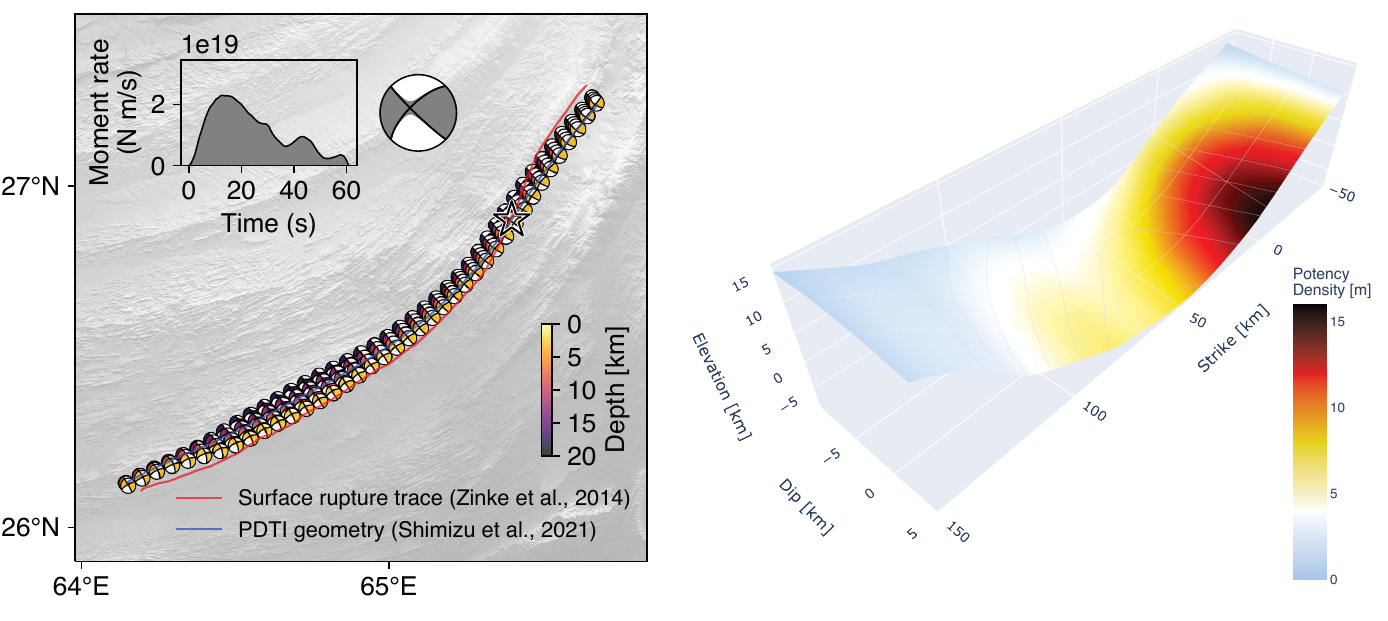}
 \caption{
 Our optimal source model ($\lambda^{-1}=0.03$, the sixth iteration). Distributed potency tensors are shown in the birds-eye view with the epicenter indicated by the star (left). The absolute potency densities are projected onto the reconstructed fault surface (right). The fault surface elevation is evaluated against a reference vertical plane with a 226-degree strike angle and the origin set at the hypocenter. The moment rate function and the moment tensor solution obtained as the integrals of the potency-rate density tensors multiplied by rigidity are also plotted. 
 }
 \label{fig:app4}
\end{figure*}

Fig.~\ref{fig:app4} presents the optimal solution thus derived ($\lambda^{-1}=0.03$, sixth iteration). The reconstructed fault forms a bowl-shaped surface with an along-strike curve that is more pronounced at shallower depths and a dip angle that become smaller at greater depths. This geometry is consistent with that expected of a listric fault. Although the input $n$-vectors corresponding to the B-spline knots of the PDTI consist of only four points along the dip, 
the reconstructed surface successfully delineates variations along both strike and dip. While these smooth strike-and-dip variations had already been captured by the beachball nodal planes without performing surface reconstruction (Fig.~\ref{fig:app2}), the three-dimensional surface reconstruction makes the interpretation significantly more intuitive. 
Along-dip elevations of a few tens of kilometers relative to the vertical reference plane are consistent with the constant-dip fault model of \citet{jolivet20142013} derived from distributed point sources. 
At the same time, similar to the alternative model of \citet{jolivet20142013} which reflects the geological setting of listric faults, the fault reconstructed in this study exhibits three-dimensional features.

While the overall geometry is consistent with the expected listric faulting and validates the developed method, specific local irregularities highlight the limitations of the current approach and define the scope for future refinements.
Since the reconstructed surface exhibits an unnatural reversal of the along-dip slope near the 150 km strike at depth (Fig.~\ref{fig:5}), the accuracy around the southwestern edge is uncertain, 
which suggests that the non-slipping zone is poorly constrained even with prior information regarding the smoothness of potency densities. 
This limitation might be overcome by trimming source faults, for example, by employing information-theoretic approaches to determine the minimum number of potency-tensor solutions required to describe source faults~\citep{thurin2024sparse}. 
Similar approaches might objectively determine the number of faults, which was manually decided to be one in this study. 
The elevation north of the epicenter is estimated systematically lower than the surface trace as in \citet{shimizu2021construction}, likely an artifact resulting from smoothing out a near-hypocenter stepover visible in the surface trace. 
Incorporating surface rupture traces as an a priori constraint would bring fault shapes closer to reality.
Although method validation remains in progress as above, anticipating that ensuing case studies will reveal the workings of the fault reconstruction from potency, we present the result of the simplest method (eq.~\ref{eq:iteration_3dfaultinfPDTI}) as a baseline solution.

\section{Conclusion}
We have investigated surface reconstruction from the density field of the potency tensor, where a smooth surface is constructed from the $n$-vector field indicated by the nodal planes of distributed potency-density-tensor solutions. 
The practical contribution of this work is the derivation of a formula that enables the three-dimensional surface reconstruction unaddressed by \citet{shimizu2021construction}. 
A key finding is that the problem of earthquake fault reconstruction is overdetermined, thus necessitating a probabilistic, Bayesian approach: a criterion is necessary to reduce the two degrees of freedom of the surface $n$-vector to the single degree of freedom of the surface elevation $z$. 
This overdetermination is a byproduct of the potency's capacity to capture the inelastic earthquake source beyond displacement discontinuity. 
The $n\mapsto z$ mapping is a dimensionality reduction, entrusting the definition of what constitutes faulting to our subjectivity to discern the true nature of the earthquake fault.
We have introduced the prior that treats an earthquake fault as a smooth interface embedded in a three-dimensional space, compatible with a dislocation within a linear continuum, which will provide one normal form as in slip inversion.
The same principle applies to surface reconstruction from distributed point sources, and our formulation is directly applicable to such cases. 

Because the degrees of freedom of the source model at each coordinate are six for inelastic strain and five for displacement discontinuity at most, the PDTI estimating the potency density is higher-dimensional than slip inversion estimating the displacement discontinuity. The underlying physics is eq.~(\ref{eq:defofnvector_PDTI}), where inelastic strain is represented as the tensor product of two vectors. Equation~(\ref{eq:defofnvector_PDTI}) signifies the requirement for volume-less areal source distributions, serving as a constraint when projecting inelastic strain onto displacement discontinuity. While this constraint is a simple determinant-free condition for point sources, it raises the regularity condition of the surface indicated by the $n$-vector field (eq.~\ref{eq:commutability_constraint_nvect}) for areal sources. The well-established fact in point-source analysis that a seismic source cannot be exhaustively expressed as a displacement discontinuity has been largely missed in finite-source analyses since the advent of slip inversion. However, this reappears in the PDTI as the overdetermined nature of surface reconstruction, where the areal source cannot be fully described by a displacement discontinuity. The optimal model in the PDTI is generally not a displacement discontinuity even after the earthquake fault is reconstructed. A research field of continuum theory originating with seismological descriptions of point sources, the inference of inelastic strains with excess degrees of freedom is thus formalized as the PDTI for areal sources. Our reconstruction of a smooth surface is but the first step toward decoding the volumetric rupture episodes compressed into the excess degrees of freedom in the areal source representation.

Having elucidated the above essence of surface reconstruction, we have formulated and solved the problem of stochastic surface reconstruction from a given noisy $n$-vector field. 
Our synthetic tests verify that the derived solution successfully reproduces the original surface from noisy $n$-vector data. 
Even for narrow faults, our three-dimensional probabilistic approach can better reproduce the original shape than previously proposed deterministic quasi-two-dimensional approaches. 
A preliminary analysis of the 2013 Balochistan earthquake captures along-dip shape variability that was not resolved in \citet{shimizu2021construction}, which validates our proposal.
Neither the prior to reconstruct a fault as a smooth surface nor our recipe to perturbatively reconstruct a surface from the initial guess is yet exhaustive. Nonetheless, they establish a reference framework for earthquake fault reconstruction from the measured seismic moment.

\begin{acknowledgments}
The first author D.S. acknowledges valuable comments from Dr. Haruo Horikawa. The authors also thank Dr. Julien Thurin and an anonymous reviewer for their insightful reviews, which have substantially refined the manuscript. This work was in part supported by JSPS KAKENHI Grant Numbers JP21H05206 and JP25K01084. 
\end{acknowledgments}

\section*{Data availability}
Waveform data inverted in our application were downloaded from the Seismological Facility for the Advancement of Geoscience Data Management Center and were processed in \citet{shimizu2020development}. 
The ground topography in this paper refers to SRTMGL1~\citep{NASAJPL2013}. 

\bibliographystyle{gji}
\bibliography{PDTIbib2025}

\appendix

\section{Derivation of Eq.~(2) from the compatibility condition for areal inelastic strain fields}
\label{app:compatibility}
\renewcommand{\theequation}{\thesection.\arabic{equation}}
\setcounter{equation}{0}

Equation (\ref{eq:defofnvector_PDTI}) represents an integrability condition for inelastic strain in the limit of infinitesimal volume, which does not necessarily hold for volumetric sources even when thickness is ignored. Since the role of eq.~(\ref{eq:defofnvector_PDTI}) as the integrability condition, often played by the compatibility condition in continuum mechanics, is seldom explicitly articulated in the literature, we provide its derivation here to clarify the distinction between inelastic strain and the displacement-discontinuity representation.

The compatibility condition for a symmetric inelastic strain $\epsilon^*$ to possess a corresponding displacement field is given by~\citep{saada2013elasticity}
\begin{equation}
e_{ikr}e_{jls}\partial_i\partial_j\epsilon^*_{kl} = 0,
\end{equation}
or symbolically,
\begin{equation}
 \overrightarrow{\nabla}\times \epsilon^* \times \overleftarrow{\nabla} = 0,
\label{eq:appA2}
\end{equation}
where $e_{ikr}$ is the permutation symbol, and $\partial$ denotes partial differentiation. 

While the compatibility condition is typically imposed on the elastic strain, it violates the compatibility condition in dislocation problems. An inelastic strain must be incompatible to act as a stress source; otherwise, it could simply be compensated for by a compatible elastic deformation without generating internal stress. To preserve the compatibility of the total strain, the elastic strain absorbs this incompatibility precisely where the inelastic source exists. For an inelastic source localized as a Dirac delta function, the violation of the compatibility equation is proportional to the second spatial derivative of the delta function. Here, the overall compatibility of elastic strains is recovered when integrated three times over space. By the same token, the areal inelastic strain must also satisfy the compatibility condition in the sense of a triple spatial integral.

We apply this weakened compatibility condition to a distribution of inelastic strain with thickness $\Delta h$ along a pseudo boundary $\Gamma$. For $\boldsymbol\xi \in \Gamma$, we consider an asymptotic representation for a small $\Delta h$ while keeping the areal potency density $D := \int^{\Delta h/2}_{-\Delta h/2} dh \epsilon^* (\boldsymbol\xi + h\mathbf{n}(\boldsymbol\xi))$ constant.
In this asymptotics, the gradient in the ${\bf n}$-direction scales as $\Delta h^{-1}$. Consequently, the triple integral of eq.~(\ref{eq:appA2}) along the ${\bf n}$-direction reduces to:
\begin{equation}
{\bf n}\times D\times {\bf n}=\mathcal O(\Delta h).
\end{equation}
Finally, by evaluating the symmetric double cross product with ${\bf n}$, ${\bf n}\times l.h.s.\times {\bf n}=\mathcal O(\Delta h)$, we obtain the following weakened form of the compatibility~\citep{sato2022displacements}:
\begin{equation}
({\bf I}-{\bf n}{\bf n}^{\rm T})D ({\bf I}-{\bf n}{\bf n}^{\rm T})=\mathcal O(\Delta h),
\label{eq:compatibility_reduced}
\end{equation}
where $\bf{I}$ is the identity matrix, and ${}^{\rm T}$ denotes the transpose.

Substituting eq.~(\ref{eq:defofnvector_PDTI}) into the left-hand side of eq.~(\ref{eq:compatibility_reduced}) yields zero on the right-hand side, showing that $D$ following eq.~(\ref{eq:defofnvector_PDTI}) satisfies eq.~(\ref{eq:compatibility_reduced}) in the infinitesimal thickness limit $\Delta h \to 0$ where the small term on the right-hand side vanishes. Conversely, in the limit $\Delta h \to 0$, $D$ satisfying eq.~(\ref{eq:compatibility_reduced}) can be expressed without loss of generality as the symmetric form $D = (\tilde{D} + \tilde{D}^{\rm T})/2$ of $\tilde{D}$ such that $({\bf I} - {\bf n}{\bf n}^{\rm T})\tilde{D} = 0$, where $\tilde{D}$ can be written as ${\bf n}{\bf s}^{\rm T}$ using a vector ${\bf s}$ as specified by eq.~(\ref{eq:defofnvector_PDTI}). That is, eq.~(\ref{eq:defofnvector_PDTI}) is equivalent to eq.~(\ref{eq:compatibility_reduced}) in the limit of displacement discontinuity $\Delta h \to 0$.

Equation (\ref{eq:defofnvector_PDTI}) describes the requirement for the limit of a displacement discontinuity localized on a pseudo-boundary $\Gamma$. 
To maintain consistency with continuum mechanics, the pseudo-boundary $\Gamma$ must be a smooth surface in this limit, a geometric constraint that manifests as eq.~(\ref{eq:commutability_constraint_nvect}) in surface reconstruction. We remark that those requirements do not exclude the possibility that zero-volume inelastic strain fields, such as fault networks, may violate eqs.~(\ref{eq:defofnvector_PDTI}) and (\ref{eq:commutability_constraint_nvect}) when mapped onto a single smooth surface. A violation of these conditions implies the potential departure of the actual source geometry from the given model space.

\section{Uncertainty propagation from potency to fault shape}
\label{app:A}
\renewcommand{\theequation}{\thesection.\arabic{equation}}
\setcounter{equation}{0}
The prior of fault geometry $P(\Gamma|{\bf a})$ shown in \S\ref{sec:Gammaprior} and the posterior of potency $P({\bf a}|{\bf d})$ [denoted by $P({\bf a}|{\bf d};\boldsymbol\sigma^2,\boldsymbol\rho^2)$ in the text] determine the marginal posterior of fault geometry: 
\begin{equation}
    P(\Gamma|{\bf d})=\int d{\bf a}P(\Gamma|{\bf a})P({\bf a}|{\bf d}).
    \label{eq:defofposteriorofshape}
\end{equation}
Here, we outline the procedure to translate $P({\bf a}|{\bf d})$ into $P(\Gamma|{\bf d})$ under the condition of the dislocation limit, where the prior $P(\Gamma|{\bf a})$ concentrates on the expected normal, $n\xrightarrow{\mathcal P} n_*$ (eq.~\ref{eq:nvector_limitequality}). 

The calculation is facilitated by introducing the (marginal) posterior of the $n$-vector field $\{{\bf n_*}\}$ expected from the potency density: 
\begin{equation}
    P(\{{\bf n_*}\}|{\bf d})=\int d{\bf a}P({\bf a}|{\bf d})\int d\boldsymbol\xi\delta ({\bf n}_*(\boldsymbol\xi)-{\bf n}_*(\boldsymbol\xi;{\bf a})).
\end{equation}
The use of $P(\{{\bf n_*}\}|{\bf d})$ simplifies the functional form of the marginal posterior of fault shape. Specifically, for $P(\Gamma|{\bf a})$ given by eq.~(\ref{eq:prior_Gamma}), we obtain
\begin{equation}
    P(\Gamma|{\bf d})=\int d\{{\bf n_*}\} P(\{{\bf n_*}\}|{\bf d})\exp[\kappa\int d\boldsymbol\xi {\bf n}\cdot{\bf n_*}]/\mathcal Z.
\end{equation}
The right-hand side describes the transformation from
the posterior $P(\{{\bf n_*}\}|{\bf d})$ of the potency-implied fault normal $\{{\bf n}_*\}$ into 
the posterior $P(\Gamma|{\bf d})$ of fault shape. 
In the dislocation limit, which is achieved by $\kappa\to\infty$, 
$\exp[\kappa\int d\boldsymbol\xi {\bf n}\cdot{\bf n_*}]$ is significantly sharper than $P(\{{\bf n_*}\}|{\bf d})$ and enforces $n\xrightarrow{\mathcal P} n_*$ (eq.~\ref{eq:nvector_limitequality}), leading to 
\begin{equation}
    P(\Gamma|{\bf d})=c\int d\{{\bf n_*}\} P(\{{\bf n_*}\}|{\bf d})\int d\boldsymbol\xi \delta ({\bf n}(\boldsymbol\xi)-{\bf n}_*(\boldsymbol\xi)),
    \label{eq:priorofGamma_usingnvect1}
\end{equation}
where $c$ denotes a normalization constant. 
Equation~(\ref{eq:priorofGamma_usingnvect1}) holds for any
$P(\Gamma|{\bf a})$ satisfying $n\xrightarrow{\mathcal P} n_*$ (eq.~\ref{eq:nvector_limitequality}).

Despite its appearance, eq.~(\ref{eq:priorofGamma_usingnvect1}) does not conclude that 
the posterior $P(\Gamma|{\bf d})$ of fault surface $\Gamma$ is identical to 
the posterior $P(\{{\bf n_*}\}|{\bf d})$ of the potency-conforming $n$-vector $n_*$: 
$
    P(\Gamma|{\bf d})\neq P(\{{\bf n_*}\}|{\bf d})|_{{\bf n}_*={\bf n}}.
$
As discussed in \S\ref{sec:Gammaprior}, the $n_*$-vector field determined from potency generally exceeds the definition domain of the $n$-vector field on a smooth surface $\Gamma$ bounded by the surface slope integrability (eq.~\ref{eq:commutability_constraint_nvect}). 
Consequently, eq.~(\ref{eq:priorofGamma_usingnvect1}) becomes
\begin{equation}
    P(\Gamma|{\bf d})= cP(\{{\bf n_*}\}|{\bf d})|_{{\bf n}_*={\bf n}\cap \partial_x(n_y/n_z)=\partial_y(n_x/n_z)}.
    \label{eq:posteriorofshape}
\end{equation}
Thus, searching for the optimum $n$-vector field of a smooth surface from the posterior of potency is a conditional optimization constrained by $\partial_x(n_y/n_z)=\partial_y(n_x/n_z)$. The inference of the optimum elevation as done in \S\ref{sec:Gammaprior} automatically satisfies $\partial_x(n_y/n_z)=\partial_y(n_x/n_z)$, thereby evading the complexity of explicitly handling the constraint. 
The surface reconstruction from this $P(\Gamma|{\bf d})$ is a nonlinear analysis based on eq.~(\ref{eq:nvectorfromslopes}), which relates the surface normal ${\bf n}$ with the surface slope $\nabla z$. 
The evaluation of eq.~(\ref{eq:nvectorfromslopes}) follows a similar procedure to that shown in \S\ref{subsec:solvingSPDE_elevation}.

We note that the non-uniqueness of functional forms of $P(\Gamma|{\bf a})$, which was an issue in surface reconstruction from given potency, vanishes in surface reconstruction from the posterior of potency $P(\Gamma|{\bf d})$ (eq.~\ref{eq:posteriorofshape}). 
This observation offers insight into why the overdeterminacy of surface reconstruction from potency has not posed any problems in slip inversion in inferring fault geometry. 
After all, we must resign ourselves to one of two forms of subjectivity: the subjectivity inherent in projecting observed inelastic strains onto displacement discontinuities, $P({\bf a},\Gamma|{\bf d})=P({\bf a}|{\bf d})P(\Gamma|{\bf a},{\bf d})\to P({\bf a}|{\bf d})P(\Gamma|{\bf a})$ in the text, or the subjectivity premised in slip inversion that forbids all earthquake sources except displacement discontinuities $P({\bf a},\Gamma|{\bf d})=P({\bf a}|\Gamma,{\bf d})P(\Gamma|{\bf d})$ (\S\ref{sec:probabilisticShimizu2021}).

\section{Correction of $\Gamma_*$ in evaluating non-integrable potency components}
\label{app:B}
\renewcommand{\theequation}{\thesection.\arabic{equation}}
\setcounter{equation}{0}

The evaluation of $\Gamma_*$ in the text (\S\ref{subsec:solvingSPDE_elevation}) assumed negligible non-integrable components of the $n$-vector $n_*$ expected from potency, since the subjectively determined functional form of the prior $P(\Gamma|{\bf a})$ matters for finite non-integrable components
(\S\ref{sec:Gammaprior}). 
To evaluate that dependence, we here present the evaluation of $\Gamma_*$ for given $P(\Gamma|{\bf a})$ when evaluating the non-integrability of $n_*$. 
We use eq.~(\ref{eq:prior_Gamma}) with $\kappa\to\infty$ as a representative case. The same procedure applies to other forms of $P(\Gamma|{\bf a})$.

Even in the dislocation limit 
$n\xrightarrow{\mathcal P} n_*$ (eq.~\ref{eq:nvector_limitequality}, $\kappa\to\infty$ for eq.~\ref{eq:prior_Gamma}), the $n$-vector of the reconstructed surface inevitably deviates from the expectation $n_*$ of potency if the non-integrable components of $n_*$ exist. 
In this limit, what becomes infinitesimal is the fluctuation of $n$ from $n_*$ with non-integrable components removed, that is, the $n$-vector of $\Gamma_*$ itself, denoted by $n_*^\prime$: 
\begin{equation}
    \lim_{\kappa\to\infty}{\bf n}(\boldsymbol\xi)={\bf n}_*^\prime(\boldsymbol\xi;\Gamma).
\end{equation}
If the non-integrability of $n_*^\prime$ is not negligible, the fast convergent series in the dislocation limit is the Taylor series of ${\bf n}$ around ${\bf n}_*^\prime$, not the Taylor series of ${\bf n}$ around ${\bf n}_*$ conducted in the text.

We now perform this corrected series expansion to account for non-integrable $n_*$ components dropped in \S\ref{subsec:solvingSPDE_elevation}. 
Using the derivative of the basis vectors
\begin{equation}
\left(
    \begin{array}{c}
         \frac{\partial}{\partial\theta}  \\
          \frac{\partial}{\partial\phi}
    \end{array}
    \right)
    \left(
    {\bf n}\,\,{\bf e}_\theta\,\,{\bf e}_\phi
    \right)=
    \left(
    \begin{array}{ccc}
         {\bf e}_\theta& -{\bf n} & {\bf 0}\\
         {\bf e}_\phi\sin\theta& {\bf e}_\phi\cos\theta &-{\bf n}\sin\theta-{\bf e}_\theta\cos\theta
    \end{array}
    \right),
\end{equation}
we obtain the second-order representation of ${\bf n}$ around ${\bf n}_*^\prime$ with respect to the angle fluctuations $\delta\Omega^\prime$ from ${\bf n}_*^\prime$: 
\begin{equation}
\begin{aligned}    
    {\bf n}=\left[1-\frac 1 2 (\theta-\theta_*^\prime)^2-\frac{\sin^2\theta_*^\prime}2(\phi-\phi_*^\prime)^2\right]{\bf n}_*^\prime
    \\
    +\left[(\theta-\theta_*^\prime)-\frac  12 \sin\theta_*^\prime\cos\theta_*^\prime(\phi-\phi_*^\prime)^2\right]{\bf e}_\theta^{*\prime}
    \\
    +\left[\sin\theta_*^\prime(\phi-\phi_*^\prime)+\cos\theta_*^\prime(\theta-\theta_*^\prime)(\phi-\phi_*^\prime)\right]{\bf e}_\phi^{*\prime}.
    \end{aligned}
    \label{eq:n_exactseries}
\end{equation}
Here, $(\theta_*^\prime,\phi_*^\prime)$ denotes the position of ${\bf n}_*^\prime$ on a unit hemisphere, 
and ${\bf e}_\theta^{*\prime}$ and ${\bf e}_\phi^{*\prime}$ represent associated basis vectors. Note that cubic and higher orders become infinitesimally smaller in the dislocation limit. 
Especially for 
$P(\Gamma|{\bf a})$ of eq.~(\ref{eq:prior_Gamma}) with $\kappa\to\infty$, the inner product between ${\bf n}_*$ and ${\bf n}$ expands as
\begin{flalign}
    {\bf n}_*\cdot{\bf n}=&{\bf n}_*^\prime\cdot{\bf n}-({\bf n}_*^\prime-{\bf n}_*)\cdot{\bf n}
    \\
    =&\left[1-\frac 1 2 (\theta-\theta_*^\prime)^2-\frac{\sin^2\theta_*^\prime}2(\phi-\phi_*^\prime)^2\right]-({\bf n}_*^\prime-{\bf n}_*)\cdot{\bf n}.
\end{flalign}
The first term mirrors the right-hand side of eq.~(\ref{eq:nnstarinnerproduct}) with $(\theta_*,\phi_*)$ replaced by $(\theta_*^\prime,\phi_*^\prime)$, functioning analogously to the terms derived in the main text. 
The second term represents the correction arising from the deviation of ${\bf n}_*^\prime$ from ${\bf n}_*$, explicitly written as
\begin{equation}
\begin{aligned}    
    ({\bf n}_*^\prime-{\bf n}_*)\cdot{\bf n}=\left[1-\frac 1 2 (\theta-\theta_*^\prime)^2-\frac{\sin^2\theta_*^\prime}2(\phi-\phi_*^\prime)^2\right](1-{\bf n}_*\cdot{\bf n}_*^\prime)
    \\
    -\left[(\theta-\theta_*^\prime)-\frac  12 \sin\theta_*^\prime\cos\theta_*^\prime(\phi-\phi_*^\prime)^2\right]({\bf n}_*\cdot{\bf e}_\theta^{*\prime})
    \\
    -\left[\sin\theta_*^\prime(\phi-\phi_*^\prime)+\cos\theta_*^\prime(\theta-\theta_*^\prime)(\phi-\phi_*^\prime)\right]({\bf n}_*\cdot{\bf e}_\phi^{*\prime}).
    \end{aligned}
\end{equation}
The angular fluctuations $\theta-\theta_*^\prime$ and $\phi-\phi_*^\prime$ are given by eqs.~(\ref{eq:deltatheta}) and (\ref{eq:deltaphi}) in the text with the substitution $(\theta_*,\phi_*)\to(\theta_*^\prime,\phi_*^\prime)$:
\begin{flalign}
    \theta-\theta_*^\prime&= -n_{*z}^{\prime2}( n_{*\parallel x}^\prime z_{,x}+n_{*\parallel y}^\prime z_{,y}+n_{*\perp}^\prime/n_{*z}^\prime)
    \label{eq:thetaflucutaion_exact}
    \\
    \phi-\phi_*^\prime&= -n_{*\perp}^{\prime-1} n_{*z}^\prime(n_{*\parallel x}^\prime z_{,y}- n_{*\parallel y}^\prime z_{,x}). 
    \label{eq:phiflucutaion_exact}
\end{flalign}
The remaining derivation follows the same logic as in the main text. The final result for the optimal surface $\Gamma_*$ is: 
\begin{equation}
    \begin{aligned}
    \Gamma_*={\rm argmin}
    \int d\Sigma\{
    [(\theta-\theta_*^\prime)^2+\sin^2\theta_*^\prime(\phi-\phi_*^\prime)^2]
    ({\bf n}_*\cdot{\bf n}_*^\prime)
    \\+[-2(\theta-\theta_*^\prime)+ \sin\theta_*^\prime\cos\theta_*^\prime(\phi-\phi_*^\prime)^2]({\bf n}_*\cdot{\bf e}_\theta^{*\prime})
    \\
    -2[\sin\theta_*^\prime(\phi-\phi_*^\prime)+\cos\theta_*^\prime(\theta-\theta_*^\prime)(\phi-\phi_*^\prime)]({\bf n}_*\cdot{\bf e}_\phi^{*\prime})\}.
    \end{aligned}
    \label{eq:Gammastarcorrected}
\end{equation}
Equation~(\ref{eq:Gammastarcorrected}) is a self-consistent equation, where the condition to determine $\Gamma_*$ contains 
the $n$-vector of $\Gamma_*$ (${\bf n}_*^\prime$) and the basis vectors
(${\bf e}_\theta^{*\prime}$ and ${\bf e}_\phi^{*\prime}$) at the hemispherical position $(\theta_*^\prime,\phi_*^\prime)$ of that $n$-vector, which is recursively solvable. 
For effectively integrable $n_*$, 
${\bf n}_*^\prime \simeq {\bf n}_*$ holds, reducing eq.~(\ref{eq:Gammastarcorrected}) to eq.~(\ref{eq:optimaloptimal_explicit}) in the main text. 
Corrections for forms $P(\Gamma|{\bf a})$ other than eq.~(\ref{eq:prior_Gamma}) can be similarly evaluated using eqs.~(\ref{eq:n_exactseries}), (\ref{eq:thetaflucutaion_exact}), and (\ref{eq:phiflucutaion_exact}). 

\label{lastpage}
\end{document}